# Experimental evolution of diverse *Escherichia coli* metabolic mutants identifies genetic loci for convergent adaptation of growth rate

Thomas P. Wytock[1¶], Aretha Fiebig[2¶], Jonathan W. Willett[2], Julien Herrou[2], Aleksandra Fergin[2], Adilson E. Motter[1,3*], Sean Crosson[2,4*]

[1] Department of Physics and Astronomy, Northwestern University, Evanston, Illinois, United States of America
[2] Department of Biochemistry and Molecular Biology, University of Chicago, Chicago, Illinois, United States of America
[3] Northwestern Institute on Complex Systems, Northwestern University, Evanston, Illinois, United States of America
[4] Department of Microbiology, University of Chicago, Chicago, Illinois, United States of America
* Corresponding Author
E-mail: motter@northwestern.edu, scrosson@uchicago.edu
¶These authors contributed equally to this work

## Abstract

Cell growth is determined by substrate availability and the cell's metabolic capacity to assimilate substrates into building blocks. Metabolic genes that determine growth rate may interact synergistically or antagonistically, and can accelerate or slow growth, depending on the genetic background and environmental conditions. We evolved a diverse set of *Escherichia coli* single-gene deletion mutants with a spectrum of growth rates and identified mutations that generally increase growth rate. Despite the metabolic differences between parent strains, mutations that enhanced growth largely mapped to the core transcription machinery, including the β and β' subunits of RNA polymerase (RNAP) and the transcription elongation factor, NusA. The structural segments of RNAP that determine enhanced growth have been previously implicated in antibiotic resistance and in the control of transcription elongation and pausing. We further developed a computational framework to characterize how the transcriptional changes that occur upon acquisition of these mutations affect growth rate across strains. Our experimental and computational results provide evidence for cases in which RNAP mutations shift the competitive balance between active transcription and gene silencing. This study demonstrates that mutations in specific regions of RNAP are a convergent adaptive solution that can enhance the growth rate of cells from distinct metabolic states.

## Author summary

The loss of a metabolic function caused by gene deletion can be compensated, in certain cases, by the concurrent mutation of a second gene. Whether such gene pairs share a local chemical or regulatory relationship or interact via a non-local mechanism has implications for the co-evolution of genetic changes, development of alternatives to gene therapy, and the design of combination antimicrobial therapies that select against resistance. Yet, we lack a comprehensive knowledge of adaptive responses to metabolic mutations, and our understanding of the mechanisms underlying genetic rescue remains limited. We present results of a laboratory evolution approach that has the potential to address both challenges, showing that mutations in specific regions of RNA polymerase enhance growth rates of distinct mutant strains of *Escherichia coli* with a spectrum of growth defects. Several of these adaptive mutations are deleterious when engineered directly into the original wild-type strain under alternative cultivation conditions, and thus have epistatic rescue properties when paired with the corresponding primary metabolic gene deletions. Our combination of adaptive evolution, directed genetic engineering, and mathematical analysis of transcription and growth rate distinguishes between rescue interactions that are specific or non-specific to a particular deletion. Our study further supports a model for RNA polymerase as a locus of convergent adaptive evolution from different sub-optimal metabolic starting points.

## Introduction

Single-celled organisms such as *Escherichia coli* provide excellent models to investigate the genetic basis of evolution. Forward genetic selection strategies [1, 2] combined with genome sequencing is a well-established approach to experimentally interrogate evolutionary mechanisms [3, 4]. Whereas experiments probing adaptive genetic changes of wild-type bacterial cells to a range of growth and stress conditions are prevalent [5-14], studies investigating adaptive evolutionary trajectories from different genetic starting states are less common [15-18]. We sought to identify genetic changes and corresponding gene expression changes that confer increased growth rate in five genetically distinct *E. coli* mutants with a spectrum of starting growth rates in defined glucose medium. Our study was aimed at uncovering genetic determinants of cellular growth rate involving both local epistatic interactions, and non-local epistatic interactions mediated by the metabolic network.

Defining principles of epistatic interactions is broadly important for understanding how genomes evolve [3], and has implications in the development of gene therapy approaches and the design of antimicrobials [19]. However, our understanding of the functional connectivity among genes in bacterial genomes and of how environment shapes these connections remain limited. Computational approaches have been developed to predict genetic interactions in the metabolic networks of microbial cells, and several of these interactions have been confirmed experimentally [2, 20-24]. Mathematical methods have also been developed that are directed toward explicit prediction of an extreme form of positive epistasis termed synthetic rescue – in which a deleterious mutation becomes beneficial in the presence of subsequent mutation of a gene or set of genes [22, 25] – and thereby enhances cellular growth rate [22, 26]. Reverse-genetic experimental approaches to identify negative and positive genetic interactions affecting *E. coli* growth have been reported [27, 28] and may be useful for high-throughput identification of epistatic rescue interactions. However, like other exhaustive methods, these approaches present combinatorial challenges for probing multigenic interactions.

Here, we present a proof of concept for a forward-genetic strategy to identify distinct classes of rescue interactions in *E. coli* cells cultivated on complex and chemically defined media, including demonstration of synthetic rescue interactions as predicted in [22] (S1 Fig). This adaptive evolution approach identified spontaneous mutations that enabled fast growth from multiple 'non-optimal' starting strains that harbored distinct metabolic mutations. Rescue isolates had few (only 1 or 2) genetic differences relative to the primary mutant strains, and a large majority of the rescue mutations mapped to essential genes that are required for transcription, including the β and β' subunits of RNA polymerase (RNAP) and the transcription elongation factor, NusA. To further explore this connection between transcription and growth rescue, we developed a novel computational framework to map measured changes in gene expression in these mutants to changes in cellular growth rate. Our results provide evidence for a model in which select mutations in RNAP and associated genes enhance growth from genetically and metabolically distinct starting states. However, the effects of RNAP mutations on growth rate and their interactions with the primary gene deletions depend on the chemical composition of the growth medium. Indeed, further work will be required to determine to what extent the observed mutations are driven by the genetic background versus the metabolic constraints imposed by the growth medium. Nevertheless, we conclude that mutations in distinct structural regions of RNAP are a convergent evolutionary route to enhanced growth rate.

# Results

## Evolution of increased growth rate from a wild-type genetic starting state

To provide a basis for comparison with our subsequent experiments, six independent cultures of wild-type *E. coli* strain BW25113 (WT) were serially passaged in M9 medium supplemented with 0.4 % (w/v) glucose until the culture growth rate increased (Fig 1). Two independent fast-growing colonies from each culture were sequenced, revealing mutations in *pykF* (6 of 12 strains, from 4 of 6 cultures), *rpoB* or *rpoC* (4 of 12 strains, 3 of 6 cultures), along with a series of less common mutations (Table 1). Mutations at the *pykF* [29], *rpoB* and *rpoC* [5, 6] loci are consistent with other studies of *E. coli* in minimal medium, and are directly associated with faster growth (Fig 2B). Analysis of the DNA sequencing data using the breseq package [30] demonstrated that mutations in *pykF* were a consequence of IS*5* insertions, transition mutations, and single base deletions. *rpoB* and *rpoC* mutations were a result of both transition and transversion events (Table 1).

## Rescue of slow growth from multiple genetic starting points

We next evolved strains harboring deletions of non-essential genes that cause defects in carbohydrate metabolism, amino acid metabolism, lipid and aromatic compound metabolism, phosphorus metabolism, or iron acquisition. Primary mutants were deleted for glucose-6-phosphate dehydrogenase (Δ*zwf*), polyphosphate kinase (Δ*ppk*), diaminopimelate epimerase (Δ*dapF*), isochorismate synthase 1 (Δ*entC*), or diacylglycerol kinase (Δ*dgk,* i.e. Δ*dgkA*). We validated the genetic backgrounds of all primary deletion strains by whole-genome sequencing. The choice of these strains is based on the well-documented biochemical and metabolic functions of the genes, and the fact that the deletion mutations result in minor to extensively slowed growth, relative to the WT parent strain, in minimal glucose medium (M9G). The growth rate of mutant strains was 5–70% slower than that of the WT strain, except Δ*ppk*, which grew at a statistically equivalent rate to the WT (S1 and S2 Tables). Studies examining adaptive evolution in multiple genetic backgrounds (e.g., [2]) are not prevalent in the literature, and thus a goal of this work was to apply these modern sequencing technologies to characterize the genetic and transcriptomic changes in these strains.

We conducted multiple, independent runs of adaptive evolution selecting for increased growth rate using these five primary deletion strains, serially cultivating the strains in M9G for 3–4 weeks (Fig 1). Over the cultivation period, the mean growth rate of replicate cultures increased. A fraction of each fast-growing culture was plated to isolate single colonies. Six independent measurements of logarithmic phase growth rate for these clones were conducted in M9G (Fig 2A). This approach identified suppressor (*sup*) strains that had significantly faster growth rates than their corresponding primary mutant parents. Several *sup* isolates derived from slow-growing mutants grew as fast as, or faster than the WT strain (Fig 2A, S1 Table), indicating that they had acquired mutation(s) that mitigated the slow growth phenotype of the primary gene deletion. Rescued strains were successfully isolated from all starting genetic backgrounds.

**Mutations in genes encoding RNAP subunits associated with faster growth in defined medium**

To understand the genetic basis of the observed growth phenotypes in M9G, we sequenced the genomes of the five primary single-gene deletion mutants and 16 fast-growing, adaptively evolved (AE) *sup* strains. The *sup* strains differed from their parental single-gene deletion strains at only one or two chromosomal loci, which mapped to both coding and non-coding regions of the genome (Table 1). Thirteen of the sixteen *sup* strains contained polymorphisms in the genes encoding the β (*rpoB*) or β' (*rpoC*) subunits of RNAP. One *sup* strain contained a single nucleotide polymorphism in the gene encoding the transcription termination/anti-termination factor NusA. Therefore, 14 of the 16 *sup* strains harbored changes in the sequences of either core β/β' RNAP subunits, or the essential RNAP accessory factor NusA. These mutations occurred in β and β' in regions of RNAP implicated in the control of transcription pausing and elongation (see Table 2 for references). Mutations were selected at the same position in *rpoB* or *rpoC* in multiple independent *sup* strains from different primary mutant backgrounds (e.g. *rpoC* R978, *rpoC* R1075, and *rpoB* D516; see Table 1).

In the Δ*dgk* and Δ*ppk* backgrounds, multiple strains were isolated in which the only genetic difference between the parent and the fast-growing strain was a single coding change in *rpoB*, *rpoC*, or *nusA* (Table 1). Increased growth rate can thus be directly attributed to a single amino acid difference in either the RNAP core enzyme or in NusA. Mutations at shared positions in fast-growing strains arose from genetic backgrounds in which growth rates varied significantly. Single amino acid changes near the RNAP jaw domain and proximal to the catalytic active site are associated with fast growth in these strains (Fig 3, Table 2). The single coding change NusA(I49N), which rescues slow growth of Δ*dgk*, is located in the N-terminal domain of the protein, which interacts directly with the RNA exit channel region of RNAP [42]. Again, we applied the breseq analysis approach [30] to account for possible IS element-mediated gene inactivation in our *sup* strains. In the case of *zwf-sup*1 and *zwf-sup*2, we uncovered evidence for IS*5* insertion junctions in *clsA* and *cydA*, respectively. These mutations are in addition to lesions in *rpoC* and *rpoB*. Mutations in the remaining *sup* strains were a result of transition, transversion, or indel mutations (Table 1).

**Mutations in *sup* strains generally reduce growth rate in complex medium**

Given the consistent emergence of mutations that increased growth rate in M9G across multiple primary mutant backgrounds, we considered whether these same mutations would increase growth rate in a chemically distinct medium. Therefore, we measured growth rates of the WT strain, primary mutants, and the fast-growing *sup* strains in LB. While *sup* strains grew as fast as, or faster than WT in M9G, these strains generally grew slower than WT in LB, with the exception of Δ*dgk-sup*3, Δ*entC-sup*2, Δ*entC-sup*3, and the Δ*ppk-sup* mutants, which were statistically indistinguishable from WT in LB (Fig 4, S2 Table).

We conclude, as have others [5], that the physiological changes resulting from *sup* mutations provide a condition-specific growth advantage in the defined glucose medium in which these mutations were selected. These genetic changes do not confer increased fitness in a medium with a fundamentally different composition.

**Identification of synthetic rescue interactions using a genetic "knock-in" approach**

Synthetic rescue is a form of epistasis in which a growth defect caused by one deleterious mutation can be mitigated by perturbation of another gene, or set of genes, which are individually non-beneficial on its (or their) own [22]. A simple form of synthetic rescue involves a local chemical or regulatory connection. For example, in *Escherichia coli* grown on gluconate, deletion of KDPG aldolase (*eda*) results in accumulation of a toxic metabolic intermediate; loss of phosphogluconate dehydratase (*edd*) prevents accumulation of this intermediate and thus the deletion of *edd* rescues the growth defect caused by deletion of *eda* [43, 44]. Rescue epistasis can also result from non-local interactions mediated by the metabolic network when a suboptimal response to a deleterious primary mutation is brought closer to the optimum by further mutations, which act as rescue mutations, even though they are non-beneficial on their own [26, 45]. We illustrate the conditions necessary for synthetic rescue and how they relate to our experiments in S1 Fig.

To test whether the fast growth observed for a given *sup* strain in our mutant set required absence of the primary metabolic gene, we restored the locus of each deleted metabolic gene with a WT copy in each *sup* strain. We measured growth rates of these "knock-in" strains in M9G. Restoration of the primary deletion site to WT yielded strains that we term "restore-*sup*", which all grew faster than WT in M9G (Fig 2A). Furthermore, the growth rates of the restore-*sup* strains were not statistically distinct from the *sup* strains, though *dapF* and *entC* restore-*sup* trended toward increased growth rate over the Δ*dapF*-*sup* and Δ*entC*-*sup* strains. These results provide evidence that, in contrast with the case of *edd/eda* [44] outlined above, the growth phenotypes conferred by the *sup* alleles are not manifested through an interaction with the primary deleted gene. Rather, we conclude that the *sup* mutations that confer faster growth in M9G are independent of whether the primary genetic lesion is present. This fact leaves open the possibility that *sup* mutations result from adaptation to features of minimal medium as opposed to the specific metabolic deletion. Additional experiments to evaluate the possibility that the observed mutations are specific to the cultivation conditions, such as introducing previously characterized mutations into the metabolic mutants and measuring growth, are needed to resolve this point.

Given the significant differences in *sup* strain fitness between the defined and complex media, we also measured growth of the restore-*sup* strains in LB. In the case of Δ*ppk*, Δ*entC*, and Δ*dgk* strains, restoration of the primary genetic lesion to the WT allele had no effect on growth rate. The presence of the *sup* mutations in either genetic background results in a WT growth rate. The Δ*zwf*-*sup*2 strain provides an interesting case in which the *sup*2 mutations *rpoB*(Q31R) *cydA::*IS*5* (Table 1) leads to slower growth in LB whether *zwf* is present or not (Fig 4A). The Δ*zwf*-*sup*1, -*sup*3, and -*sup*4 mutants exhibited slow growth in LB, which was restored to WT growth upon restoration of *zwf*.

Finally, we note that the Δ*dapF*-*sup*1 and -*sup*2 strains grown in LB are instances of synthetic rescue interactions (as defined, e.g., in [22] and illustrated in S1A–S1C Fig). These *sup* mutations result in significantly increased growth rates relative to Δ*dapF* alone ($p$-value < 0.0001; one-way ANOVA; Dunnett's post), though growth is not restored to WT levels (Fig 4A). Restoration of the *dapF* locus in the *dapF-restore-sup*1 and -*sup*2 strains slightly decreased fitness relative to Δ*dapF*-*sup*1 and Δ*dapF*-*sup*2, respectively. These strains all harbor the same *lrp*(T134N) allele as well as distinct mutations in the 3' end of *rpoB* (Table 1). These results show that the synthetic rescue is determined by an antagonistic interaction between the *dapF* deletion and the *lrp/rpoB* mutations that confer increased fitness in LB. Likewise, the Δ*zwf*-*sup*1 ($p$-value < 0.01) strain grown in LB exhibits synthetic rescue epistasis, illustrating the antagonistic relationship between the *zwf* deletion and the *rpoC* mutation, *clsA* mutation, or both.

Additional synthetic rescue pairs are identified for certain Δ*dapF*-, Δ*dgk*-, Δ*entC*-*sup* strains when we consider the adaptively evolved (AE) strains harboring the *sup* mutation(s) as an evolutionary starting point (Table 3). From this state, the conditions of synthetic rescue can also be met (S1F Fig). Specifically, restoring the primary deletion (Δ) is neutral and reverting the AE *sup* mutations to their WT alleles is deleterious (S1E and S1F Fig). If the *sup* mutations are reverted to their WT alleles, then restoring the primary deletion becomes beneficial.

**Transcriptomic analysis identifies expression changes associated with growth rescue**

Mutations in RNAP are by far the most common rescue solutions for the primary deletion strains assayed in this study. We also identified mutations in the transcriptional regulatory gene *lrp* in each of the fast-growing Δ*dapF-sup* strains. Given the clear connection of nearly all *sup* mutations to transcription, we hypothesized that these mutations may enable growth rescue by modulating transcription of a particular gene or set of genes. Therefore, we measured steady-state transcript levels in the wild type strain, in each of the five primary deletion mutant strains, and in a set of *sup* strains cultivated in M9G (gray background in Table 1), using RNA sequencing to define transcriptional features that are associated with fast or rescued growth.

These measurements revealed many transcriptional differences between WT *E. coli* and the five primary deletion strains: Δ*zwf*, Δ*ppk*, Δ*dapF*, Δ*entC*, and Δ*dgk.* Expression of 7 to 381 genes differed significantly between the WT strain and primary deletions (fold change > 1.5; false discovery rate adjusted *p*-value < 0.01); significant expression differences between the WT strain and primary deletions ranged from 1.7- to over 400-fold. An ontology analysis [46] of genes with dysregulated expression upon deletion of these five metabolic enzymes provides results that are congruent with the known functions of the deleted genes: DapF catalyzes the penultimate step in lysine biosynthesis [47], and the dysregulated gene set in the Δ*dapF* strain is strongly enriched ($p$-value < $10^{-4}$) in gene ontology terms associated with amino acid biosynthesis; EntC is an isochorismate synthase involved in enterobactin synthesis [48], and deletion of *entC* results in dysegulation of genes participating in enterobactin siderophore biosynthesis and transport (*p*-value < 0.05). However, we also observed large changes in transcription that are not directly related to the enzymatic functions of Zwf, Ppk, DapF, EntC, and Dgk. For example, deletion of either *dgk* or *entC* results in consistent upregulation of *flh* and *flg* genes, which are responsible for flagellar motility. This upregulation is the most significant gene expression ontology pattern in both of these datasets ($p$-value < $10^{-13}$), and has been noted in other *rpo* mutants [5], but does not occur across all *sup* strains harboring *rpoB* and *rpoC* mutations.

We envisioned two possibilities by which each gene's expression may change in *sup* strains relative to the primary mutant: 1) *restorative adaptive evolution*, in which the secondary mutations largely restore a gene's expression to WT levels, or 2) *compensatory adaptive evolution*, in which the secondary mutations move gene expression into a state that is distinct from the WT strain (Fig 5A). To assess these two possibilities for each adaptively evolved strain, we identified genes that are commonly regulated across the *sup* strains associated with each primary deletion. Such commonly regulated genes may underlie the increased growth rates in M9G that we observe in our *sup* strains.

Our transcriptome analysis approach to distinguish restorative from compensatory adaptive evolution, illustrated in Fig 5A, builds on the differential expression analysis previously discussed. We evaluated expression changes of each gene with respect to: (i) the fold change and (independently) the statistical significance of its expression as functions of the primary gene deletion; (ii) the fold change and (independently) the statistical significance of its expression as functions of the adaptively evolved *sup* mutations; and (iii) the relative sign and magnitude of these changes. Restricting our analysis to genes that exhibited a statistically significant change as a function of the primary deletion or the rescue mutations, we evaluated whether adaptive evolution restored or compensated for each gene's transcription. The magnitude and relative sign of a gene’s expression change between the deletion strain and its derived *sup* strain distinguished a restorative change from a compensatory change. Genes exhibiting completely restorative or completely compensatory responses scored 1 or -1, respectively, on a Gene Change Score (GCS) scale in which genes with insignificant changes were scored as 0. The set of identified genes may change as the thresholds for statistical significance or fold change are varied. To mitigate this possible threshold dependence, we averaged the scores calculated over a grid of the thresholds (see Materials and Methods). For a comparison with the standard differential expression, we used Venn diagrams to display the gene sets in which the acquisition of *sup* mutation(s) restores transcription to a level that is statistically indistinguishable from the WT strain.

Figure 6 summarizes the full spectrum of transcriptional changes for each *sup* strain. The GCS of each gene is shown in a histogram (Fig 6A), revealing a bias toward restorative changes over compensatory changes across the different primary deletions. The GCS analysis identifies genes with similar transcriptional profiles (GCS > 0.4 or GCS < −0.2) in all suppressors of each primary deletion background, thus refining the gene sets shown in the Venn diagram (Fig 6B).

The GCS histograms in Fig 6A reflect how similar the *sup* strains are with respect to gene expression. We compared transcriptomes of Δ*dapF*-, Δ*dgk*-, Δ*entC*-, and Δ*ppk*-*sup* strains isolated from independent adaptive evolution experiments; transcriptomes of Δ*zwf*-*sup* strains were measured for two independent colonies from the same adaptive evolution run (Table 1). As expected, RNA-sequencing samples from Δ*zwf sup*1-1 and Δ*zwf sup*1-2 were the most similar, as these strains are isogenic. The Δ*entC*-*sup* strains also had similar transcriptional profiles: all these strains carry one of two mutations, G(–58)A or G(–55)A, in the 5' leader of *menF* (a paralogous isochorismate synthase gene in *E. coli*) at a predicted translation attenuation site [49] (Table 1). These mutations result in de-repressed *menF* expression and decreased *pyrL* leader peptide, which regulates pyrimidine biosynthesis. In other cases, at least one *sup* strain had distinct transcriptional properties from the other *sup* strains derived from the same deletion parent.

Nevertheless, we observed further shared transcriptional trends across individual *sup* strains, as shown in Fig 6B. In the cases of Δ*ppk* and Δ*zwf*, where few genes have altered transcription between the WT strain and the primary deletion, approximately one-half to one-third of the genes with altered expression in the primary mutant background were statistically restored to the WT levels in the *sup* strains. The remaining three cases show a smaller fraction (15–20%) of genes restored to the WT levels. Genes with restored expression in Δ*dgk*-*sup* strains include the *liv* branched-chain amino acid transport system [50] and the molybdate and maltose transporter genes. In Δ*entC*, the *carAB* glutamine catabolism genes, *codB* cytosine transport gene, and *nadA* $NAD^+$ biosynthetic gene are restored to WT levels by the *sup*2 and *sup*3 mutations. The Δ*entC*-*sup*1 is one of two cases of growth rescue without a mutation in a component of RNAP. Finally, expression of the small RNA *spf* [51] is altered in both Δ*ppk* and Δ*zwf*, and its expression profile is reset to that of the WT strain by the *sup* mutations. The effects of small regulatory RNAs are considered in the Discussion section.

Rescue (*sup*) mutations also result in significant changes in the transcription of dozens of genes that show no response to the primary deletion before adaptive evolution (Fig 6B). RNAP mutations are generally known to have pleiotropic transcriptional effects [52]. The RNAP mutations we observe in the *sup* strains regulate genes to both adapt to M9G and bypass the detrimental effect of the primary lesion. Of the assayed primary deletion strains, the growth rate of Δ*ppk* and Δ*zwf* were nearest to that of the WT strain (Figs 2 and 4). All Δ*ppk*- and Δ*zwf*-*sup* strains share a set of genes exhibiting compensatory adaptive evolution. These shared, *sup*-specific expression changes include downregulation of the *gad* genes encoding the glutamate-dependent acid resistance system [53], and the *hdeAB* system that mitigates periplasmic acid stress [54, 55]. Both of these systems are under the control of the global regulator and nucleoid-associated protein H-NS [56, 57] and the stationary phase sigma factor, $\sigma^S$ (RpoS) [58]. We observe similar downregulation of *gad* and *hde* genes across the Δ*dapF*-*sup* strains relative to the WT strain (S2 Fig). These data provide evidence for a common underlying mechanism of growth rescue in these *sup* strains. To further explore the connection between H-NS/$\sigma^S$ dependent gene regulation and the transcriptional effects of *rpoB/rpoC sup* mutations, we performed hierarchical clustering on our RNA-sequencing data combined with publically-available transcriptomic data from *rpoS* and *hns* mutants (see S3 Table for data accession numbers). Clustering was conducted using a set of defined $\sigma^S$ and H-NS regulated genes from RegulonDB [49] (S2 Fig). The transcriptional profiles of *rpoB/rpoC* mutants cluster with those of *rpoS* deletion strains to a greater extent than by chance ($p$-value < 0.02, hypergeometric test). Expression of genes in the cluster containing *gad* and *hde* in the *hns* mutants is anti-correlated with measured expression in *rpoS* mutants and our *rpoB*/*rpoC sup* mutants (as highlighted in magenta in S2 Fig). This is consistent with an antagonistic regulatory relationship between $\sigma^S$ and H-NS at the *gad* and *hde* loci. Our results provide evidence that *rpoB*/*rpoC sup* mutations that confer fast growth in M9G alter transcription in a manner similar to that of an *rpoS* deletion mutant.

**Modeling of the ability of *sup* mutations to rescue diverse metabolic mutations**

Given the transcriptomic similarities among *sup* strains derived from Δ*ppk*, Δ*zwf,* and Δ*dapF*, we hypothesized that certain adaptive mutations have more general rescue properties. We developed a computational model to test this hypothesis *in silico*. Our computational analysis entailed 1) quantifying the transcriptional responses to *sup* mutations and 2) mapping gene expression to growth rate.

To address the first goal, we assumed that the gene expression response to *sup* mutations will be the same across genetic backgrounds. The response to *sup* mutations was assessed by taking the measured difference in gene expression between a *sup* strain and its parent deletion strain. This allowed us to estimate the gene expression that would result if *sup* mutations from one deletion strain were applied to a different deletion strain, whose growth rate can then be estimated by mapping gene expression to growth rate, which is our second goal. Toward the second goal, we acquired expression data paired with growth rates curated in [59], and supplemented it with our own RNA sequencing and growth rate data. These datasets were then used to train a k-nearest neighbors (KNN) model [60, 61] designed to convert a gene expression profile—defined by the expression levels of each gene in the genome—into a specific growth rate. Such a mapping assumes that growth rate is uniquely determined by gene expression. Briefly, as illustrated in Fig 7A, our KNN model finds the most similar gene expression profiles and computes a Euclidean distance-weighted average of the accompanying growth rate. For more details, see Materials and Methods.

We cross-validated the KNN model in two ways. In the first, and less stringent, test we validate as follows: we first simulated noisy data using the observed relationship between the mean and variance of RNA-sequencing counts (for details, see Computational modeling in the Methods). Then, we divided our dataset comprising our triplicate RNA-sequencing and growth rate measurements and the gene expression and growth rate data curated in [59] into five groups—called folds—stratified by growth rate (i. e., each fold was constrained to have the same distribution of growth rates as the entire dataset). The KNN analysis was trained on each set of four folds with combined with the simulated data, and the growth rate in the remaining fold was predicted using the folds' corresponding gene expression as input into the KNN model. Figure 7B shows that the KNN model predicted the growth rate accurately ($R^2$ = 0.84, $p$-value < $10^{-38}$), including the increase in growth rate resulting from adaptive evolution. This otherwise expected high accuracy is evidence of the correct implementation of the algorithm and the consistency of our RNA-sequencing data.

A second and more stringent test of the KNN analysis is to use only published data for training, and then predict the entirety of our measured growth rates using our gene expression data as input. The results, presented in Fig 7C, exhibit considerably more spread than Fig 7B ($R^2$ = 0.11, *p*-value < 0.015), yet still predict 35 of 57 experiments within 25% of the measured growth rate (grey shading). This number increases to 45 of 57 when we predict growth rate of the three replicates of each strain while augmenting the published training data with all our growth rate and transcriptomic measurements except for those which correspond to the replicates of the strain being predicted. The latter case is representative of the accuracy expected in our application below of the KNN analysis, while the accuracy in Fig 7C can be interpreted as a typical lower bound for out-of-sample prediction.

Having validated the KNN model, we applied it to predict whether gene expression responses to *sup* mutations measured in one primary deletion background would increase growth if applied to the other primary deletion backgrounds. Our training data consisted of the data curated in [59] plus all of our gene expression and growth rate data. For all cases in which we had RNA sequencing data, we added the transcriptional responses back to the parent deletion gene expression profiles; the resulting KNN model output is congruent with the experimental growth rate data presented in Fig 2 and S1 Table. The predicted responses of primary deletion strains to the full spectrum of rescue *sup* mutations is shown in Fig 8. The *sup* mutations that rescue growth of Δ*dapF* have limited capacity to increase growth rate of other primary mutants, while the Δ*entC-sup*3 mutations (*rpoC* R1075C and *menF* G(–58)A) increased the growth rate of all primary deletion strains. Transcriptional changes resulting from *sup* mutations in Δ*dapF*, Δ*ppk,* and Δ*zwf* have similar properties in our KNN model, as *sup* mutations derived from Δ*ppk* and Δ*zwf* suppressors rescue the Δ*dapF* strain. The same *rpoC* mutation (R1075C) was identified in Δ*dgk-sup*1 and Δ*entC-sup*3, which indicates that these *sup* mutations would rescue each other in a reciprocal manner in our KNN model. In fact, all Δ*entC*-derived *sup* mutations are predicted to rescue the Δ*dgk* parent and vice versa.

## Discussion

This study assessed multiple *E. coli* evolutionary trajectories and measured corresponding gene expression changes in strains evolved from wild-type and five distinct metabolic mutants with a spectrum of initial growth rates in minimal glucose medium. We demonstrated that mutations in particular regions of RNAP were causal in enhancing growth rate across multiple mutant backgrounds, defining regions of RNAP structure (Fig 3) involved in convergent adaptive evolution [14, 62] of growth rate in minimal glucose medium. The results presented herein build upon previous studies of wild-type *E. coli* that have cataloged RNAP mutations (among many other mutations) in fast-growing, evolved strains [5, 6, 14, 63, 64]. However, only the mutation *rpoB*(H526Y) is shared between our evolved deletion strains and these published studies. Despite these studies sharing only one mutation with our evolved deletion strains, we cannot draw firm conclusions about reproducibility of the evolutionary trajectories or whether the set of mutations we observe and those observed in previous studies would be truly distinct if saturation of adaptive evolution could be achieved in all cases.

That stated, we can still learn from the similarities our evolved WT has to previous evolved WT strains and the differences between our evolved deletion mutants and our evolved WT strains. Our evolved WT strains show frequent insertion elements and frameshift mutations disrupting *pykF*, seen previously in the evolution of WT *E. coli* B [29], in contrast with other studies of WT *E. coli* MG1655 cultivated in minimal medium [5, 6], which show prevalent *rpoB* point mutations. Our evolved deletion strains and evolved WT differ in the extent of *pykF* mutation and in the location of the *rpoB* and *rpoC* mutations. While these similarities and differences should be interpreted with caution because most of the corresponding evolution screens are not saturating, they motivate further investigation into how strongly the genetic background influences the mutations acquired during adaptation.

In addition, both the number of deletion mutants considered and the identity of observed mutations distinguish our study from others that study the adaptive evolution of slow growth deletion mutants. For example, we both evolved multiple *E. coli* deletion mutants and sequenced the resulting *sup* strains (three or more per mutant), allowing us to identify whether mutations were unique to genetic backgrounds or recur independently of them. In the limited corpus of previous work that evolved single deletion mutants with subsequent sequencing in *E. coli*, all studies focused on evolutionary trajectories arising from one deletion strain starting point per study, limiting the ability to observe similarities and differences in adaptive mutations [17, 18, 65]. Among these studies, only the adaptive evolution of Δ*pgi* yields lesions in *rpoA*, *rpoB*, and *rpoC* subunits that rescue growth [17]. However, the most common rescue mutations reported in [17] were in the stationary phase sigma factor, *rpoS*, and the NADH/NADPH transhydrogenases *udhA* and *pntAB,* which were not identified in our experiments. Adaptive evolution studies of an *rpoS* deletion mutant identified a single IS*10* insertion at the *otsB* locus in multiple independent experiments [65], which our experiments did not identify. We note that Blank and colleagues documented evolution in multiple *E. coli* mutant strains that were initially unable to grow in minimal glucose medium [18], but did not uncover *rpoB/rpoC* or *nusA* mutations as specific compensatory rescue solutions or as general lab adaptation mutations.

**Structural and functional classification of rescue mutations**

The structural and functional impact of the RNAP mutations documented in our study have not been directly tested, but four of our strains harbor mutations in the β' jaw β'i6 insertion region that may have similar phenotypic impacts to the *rpoC*(ΔV1204-R1206) mutation present in [5]. Overall, the set of mutations we identified in RNAP map to regions of core enzyme structure near the active site, secondary channel, and exit channel. Notably, the *rpoB*(D516G) substitution site observed in Δ*zwf-sup*3 and -*sup*4 matches a site that confers resistance to rifamycin in clinical isolates [66]. This residue interacts with DNA at the −4 and −5 positions relative to the transcription +1 site and may play a role in promoter escape [67]. The D959F substitution in Δ*dapF-sup*1 is in the *E. coli* lineage-specific region βi9 [31]. Further study of the growth effects of this mutation may provide insight into the function of this largely uncharacterized region of RNAP in *E. coli*. We note that the Δ*dapF-sup*1 and Δ*dapF-sup*2 mutations co-occur with identical mutant alleles of *lrp*, which has been previously identified as a mutation site in experimentally evolved strains [68].

In addition to mutations in *rpoB*, *rpoC*, and *nusA,* we identified mutations in other sites, particularly in the Δ*entC* and Δ*dapF* strains, which exhibited the slowest growth rates. All three Δ*entC-sup* strains have mutations in the *menF* promoter. This shared mutation class in independently evolved Δ*entC*-sup strains suggest that modulating metabolite flow through isochorismate synthase by changing *menF* expression is important to restore growth in cells lacking *entC*. Mutations within or upstream of the *pykF* gene, which encodes pyruvate kinase I, are associated with increased growth rate across a variety of *E. coli* genetic backgrounds [6, 29, 69, 70], and were identified in our wild-type adaptive evolution experiments (Table 1). However, *pykF* mutations are rare in our adaptively-evolved mutant strains, in which *rpoB/rpoC* are by far the most common rescue mutations. We only observe one *pykF* mutation in this set: *pykF*(T278P) in Δ*entC-sup*1 in addition to a mutation in the *menF* promoter. These two genetic changes are clearly sufficient to enhance growth in the absence of changes in *rpoB* or *rpoC*. However, Δ*entC-sup*2 and Δ*entC-sup*3 harbor mutations in *rpoB* and *rpoC*, respectively, and grow significantly faster than Δ*entC-sup*1 ($p$-value < 0.001) in M9G (Fig 2A).

We identified two distinct mutations in the *lrp* transcription factor in the three Δ*dapF-sup* strains. Our KNN analysis suggests that the transcriptional responses we measured in these strains are tailored to the Δ*dapF* background (Fig 8). In addition to genetic changes in *lrp*, we again observed mutations in *rpoB* in Δ*dapF-sup*1 and Δ*dapF-sup*2. Δ*dapF-sup*3 does not have an RNAP-associated mutation. Rather, we identified a mutation 40bp upstream of *pyrE* in the promoter region between *pyrE* and *rph* in this strain. This intergenic mutation may disrupt the putative *rph* terminator [49], potentially increasing readthrough of the *rph-pyrE* transcript and enhancing pyrimidine synthesis. The presence of *lrp* mutations in all three Δ*dapF-sup* speaks to the importance of these mutations in growth rescue of this strain.

**Features of gene expression**

Gene expression analysis of the primary mutants and the derived *sup* strains provides some insight into the mechanism by which the chromosomal mutations enhance growth rate. However, there are notable changes in transcription across strains that remain challenging to link to growth rate. For example, transcriptional modulation of small regulatory RNAs is prominent in several RNA-sequencing datasets in our study (Fig 6B). The small RNA spot 42 (*spf*) has reduced expression in the Δ*ppk* strain and is restored to WT levels in *sup* strains. In addition, *chiX* is specifically upregulated in the Δ*dgk-sup* strains, *isrC* is upregulated in Δ*dapF* and partially restored in *sup* strains, and *arrS* and *cyaR* are both specifically dysregulated in the Δ*ppk*- and Δ*zwf-sup* strains. We further observed a pronounced upregulation of the *flg* and *flh* flagellar genes in the Δ*entC*- and Δ*dgk-sup* strains. This response may be related to a reported inverse relationship between growth potential of the environment and motility [71].

We note a common downregulation of *gad* and *hde* genes across the Δ*dapF*-, Δ*entC-,* Δ*ppk-,* and Δ*zwf-sup* strains (S2 Fig, magenta text). Both *gad* [56, 72] and *hde* [57, 73] are well-described targets of the nucleoid-like protein H-NS, which functions as a global regulator of gene expression and a determinant of chromosomal structure [74]. These genes are typically silenced by H-NS during logarithmic growth, and a number of identified RNAP rescue mutations shift *gad* and *hde* toward a less expressed state. Regulation of *gad* and *hde* may involve the stationary phase sigma factor ($\sigma^S$), which is known to function in concert with H-NS at these (and other) promoters [75, 76]. A possible role for H-NS and/or $\sigma^S$ in *sup* strains carrying RNAP mutations is also congruent with the observed restoration of *flg*, *flh*, and *fli* flagellar gene expression across multiple primary deletion backgrounds (Δ*dgk,* Δ*entC,* and Δ*ppk*) (see Fig 6B). H-NS-dependent activation of flagellar gene expression in *E. coli* is established [77-80]. Moreover, deletion of *rpoS* is known to enhance motility [81, 82] via activation of flagellar gene expression when cells are grown in M9G [83]. Again, these gene expression data support a model in which *rpoB*/*rpoC* fast-growing *sup* mutants have transcriptional features of an *rpoS* mutant, as outlined in the Results section (S2 Fig). These data provide evidence, albeit correlative, that transcriptional changes mediated by RNAP rescue mutations may involve a shifted competitive balance between nucleoid associated proteins like H-NS and RNAP loaded with the primary sigma factor $\sigma^{70}$ or stationary factor, $\sigma^S$. Competition between these factors at *gad* loci has been previously discussed [72].

**Mapping gene expression to growth rate**

The KNN model parses how the same global transcriptional responses to *sup* mutations vary in their ability to rescue growth, depending on whether or not they are implemented in the genetic background in which they were selected. Figure 8A may be read as a matrix, in which each subplot is a row indicating the genetic background and each alternating shading forms a column encapsulating the set of *sup* mutations derived in that background. Some *sup* mutations are predicted to increase growth in multiple deletion backgrounds. For example, the Δ*dgk-sup* mutations (column 2) rescue the Δ*entC* deletion strain (row 3) and Δ*entC-sup* mutations (column 3) reciprocate, rescuing the Δ*dgk* strain (row 2).. However, other mutations may have divergent growth impacts, despite exhibiting similarities in transcriptional responses. As an example, consider the states formed by the combinations of the Δ*dapF*, Δ*ppk*, and Δ*zwf* backgrounds and their derived *sup* mutations (rows and columns 1, 4, and 5), we note an interesting predicted hierarchy of rescues in which: 1) the Δ*zwf-sup* mutations increase growth rate of Δ*dapF*, Δ*ppk*, and Δ*zwf*; 2) the Δ*ppk-sup* mutations increase growth rate of Δ*dapF* and Δ*ppk*; and 3) the Δ*dapF-sup* mutations increase the growth rate of only Δ*dapF*. The origin of this non-reciprocity in genetic interactions is not yet understood and is worthy of further study.

**Toward systematic discovery of synthetic rescue interactions**

Table 3 lists the seven examples of synthetic rescue mutation combinations satisfying the conditions laid out in S1C and S1F Figs. Cultivation of Δ*zwf-sup*1, Δ*dapF-sup*1, and Δ*dapF-sup*2 strains in LB subsequent to selection in M9G revealed synthetic rescue interactions (Fig 4A and S2 Table) – a proof-of-concept that our protocol can identify synthetic rescue epistasis. It is likely that suboptimal growth of the WT strain in M9G confounded discovery of additional forward synthetic rescue interactions using this approach. While it has been previously noted that mutations acquired during evolution in defined medium are typically maladaptive in complex medium [5], not all mutations we acquired in defined medium were deleterious in complex medium (Figs 2 and 4). Using classes of relationships between adaptive mutation and growth differences in defined and complex medium, an *a posteriori* interpretation of our experiments identifies additional synthetic rescue interactions but with the Δ*dapF-sup*2, Δ*dgk-sup*1, Δ*dgk-sup*2 and Δ*entC-sup*3 strains as starting points and the removal of *sup* mutations and restoration of deletions as the synthetic rescue interaction pair (S1E Fig). In future research, one may consider directly implementing these ordered pairs experimentally in pursuit of uncovering the mechanisms driving rescue and, in particular, determine whether non-genetic factors contribute to the effect.

It is instructive to contextualize our synthetic rescue findings in the broader literature. While our approach to find synthetic rescues is the first attempt in *E. coli*, others have implemented a similar approach in the eukaryote, *Saccharomyces cerevisiae* [84]. This study includes the key experimental components of our method: adaptive evolution of single gene knockouts including growth rate measurements, gene sequencing of suppressor strains, and transcriptomics (although microarray instead of RNA-sequencing). Screens for synthetic *lethal* interactions have been carried out in *E. coli* [27, 28], but no comprehensive study of synthetic *rescue* interactions has been reported for this or other bacteria. Synthetic rescues are harder to detect than synthetic lethal interactions because they require more than a (binary) classification into growth and no-growth states. Nevertheless, large-scale studies in *Saccharomyces cerevisiae* have systematically explored pairwise suppressive epistasis [25, 85, 86]. Our approach to find synthetic rescue interactions differs from that body of work in several respects. First, mutations in protein or mRNA degradation machinery were commonly observed to suppress primary growth defects in yeast, whereas in *E. coli* we most frequently observed mutations in the core transcription machinery. Both mechanisms have the potential to globally remodel the metabolic capacity of a cell with limited overall genetic changes. Second, while high throughput studies have the advantage of being comprehensive, they require sifting through non-functional "rider" mutations [86], making true rescue mutations difficult to identify and leaving interactions undiscovered (as evidenced by a comparison with a targeted study [87]). When contrasted with large-scale studies in yeast, our small-scale forward genetic study in *E. coli* offers higher sensitivity, enhanced targeting of specific genes, and broader applicability. Finally, our analysis goes beyond pairwise interactions, and takes advantage of adaptive laboratory evolution to uncover synthetic rescue genetic combinations in which the rescue partners are not necessarily gene deletions.

Synthetic rescues are more than a mere counterintuitive form of genetic interaction: they highlight genetic systems when targeted in combination can select against antibiotic resistance. Using each set of gene mutations in a synthetic rescue pair as a target for an antibiotic drug, the resulting pair of antibiotics will exhibit a suppressive interaction [19]. That is, a combination of the two drugs can be weaker than one of the two drugs alone. Thus, as shown in [88], the drug combination can select against cells that have developed resistance to the suppressor drug in the pair. Given the central role of RNAP in the synthetic rescue interactions identified in our study, and that the fact that drugs targeting RNAP are already approved for clinical use, it is reasonable to consider the development of antibiotic drug pairs with one targeting RNAP. Such a network-minded strategy opens the door for the re-purposing of extant pharmaceuticals to combat antibiotic resistant bacterial pathogens.

## Materials and Methods

### Experimental procedures

**Genetic manipulation, strain cultivation, and growth rate determination.** Strains used in this study are included in S1 Table. For routine growth, mutants were grown on either Luria-Bertani Broth (LB) or LB agar plates supplemented with the appropriate antibiotics at the following concentrations when required; kanamycin 50 µg/mL or ampicillin 100 µg/mL. All *E. coli* mutants were obtained from the *E. coli* Keio mutant collection [89]. The WT allele of each primary mutation was restored to the fast-growing suppressor background using a double-crossover recombination strategy that employed the plasmid pKOV (Addgene plasmid # 25769) [90]. Primary integrants were counter-selected in the presence of 10% sucrose weight/vol (w/v) on LB agar containing no sodium chloride. The genetic identity of all mutants was verified by PCR and sequencing.

To measure growth rates, overnight cultures were first started from freshly grown colonies in 2mL M9 with 0.4% (w/v) of glucose (M9G) or 2mL LB and shaken at 37˚C overnight. The next day, triplicate cultures were diluted to initial optical density at 600nm ($OD_{600}$) of 0.01 in 2mL of growth medium in a 13x100 mm borosilicate tube, and they were allowed to grow at 37°C inclined at a 45° angle at 200 rpm in an Infors Shaker. The $OD_{600}$ was measured every 20 minutes for LB and every 40 minutes for M9G for six independent cultures. These data included the log-linear region of the growth curve, which were fit to the exponential growth equation $OD_{600}(t)= e^{kt}$, where $k$ is growth rate (1/h) and $t$ is time (h). The mean and standard deviation were calculated from the rate values calculated from each of the six cultures (Figs 2 and 4; S1 and S2 Tables). A flow chart of our rescue selection approach is outlined in Fig 1.

**Selection for spontaneous fast-growing *sup* mutants.** Adaptive laboratory evolution experiments to identify spontaneous mutations that rescued slow growth of *ppk::kan*$_R$ (Δ*ppk*), *zwf::kan*$_R$ (Δ*zwf*), *entC::kan*$_R$ (Δ*entC*), *dgk::kan*$_R$ (Δ*dgk*), and *dapF::kan*$_R$ (Δ*dapF*) strains were carried out in M9G. Three independent 5mL cultures in 20x150 mm borosilicate tubes were serially passaged for either 21 days (>230 doublings for Δ*entC* and >450 for Δ*dgk*) or 28 days (>330 doublings for Δ*ppk,* >260 for Δ*zwf,* and >210 for Δ*dapF*), shaking at 37°C inclined at a 45° angle at 200 rpm in an Infors Shaker, to select for spontaneous mutants that have increased rate of growth (S1 Table). Cultures were serially passaged two times daily to maintain growth in prolonged exponential phase. Before passaging, the $OD_{600}$ was measured and used to determine the volume of culture needed for passaging. By the end of passaging, cultures were clearly growing at a faster rate than when we started the experiment. On the final day, these fast-growing cultures were streaked onto LB agar to isolate single clones. Three single colonies from each selection were picked, saved and used for growth rate measurements to confirm that individual isolates grew faster than the primary single deletion strain from which they were derived. A similar strategy was employed to identify select for spontaneous mutants in the wild-type strain with faster growth in M9G than the parental strain. In this case, six independent selections were set up in parallel. Each selection was passaged by a 1:100 dilution every 12 hours for 12 days at which point the bulk culture exhibited faster growth than the parent. After 24 passages, each culture was streaked onto LB agar to isolate single clones. Two colonies were chosen from each selection. Growth rates were measured as above for each clone in M9G and in LB in four independent cultures over 2 days.

**Mapping adaptive mutations by whole-genome sequencing.** We isolated genomic DNA from *E. coli* primary deletion and *sup* strains, and from WT and the derived *fast* mutant strains using a standard guanidinium thiocyanate extraction and isopropanol/ethanol precipitation. The DNA was randomly sheared and libraries were prepared for whole-genome shotgun sequencing using an Illumina HighSeq 2500 (50-bp single end reads). The whole-genome sequence data from each strain was assembled to the *E. coli* BW25113 WT genome template [91], and polymorphisms were identified using Geneious v 7.1.4 [92]. Additionally, we reanalyzed our genome sequencing data using the *breseq* analysis pipeline to account for possible IS element insertions [30]. Breseq confirmed the single nucleotide polymorphisms (SNPs) and small insertion/deletions identified by Geneious, and revealed some cases of insertion of mobile insertion sequences (IS elements). Each sequenced genome library yielded an average of 8.4 million reads, resulting in average depth of coverage greater than 125x. Individual sequencing reads for each primary mutant strain and corresponding *sup* strains have been deposited in the NCBI Sequence Read Archive under the submission number SUB1785225 (BioProject PRJNA339661).

**Measuring steady-state transcript levels by RNA sequencing.** We isolated RNA for sequencing from each primary mutant strain and multiple independent *sup* strains (Δ*dapF-sup*1, *-sup*3; Δ*dgk-sup*1, *-sup*2, *-sup*3; Δ*entC-sup*1, *-sup*2, *-sup*3; Δ*ppk-sup*1, *-sup*2, *-sup*3; two isogenic colonies of Δ*zwf-sup*1) in triplicate across three separate days. Overnight cultures of each strain were cultivated in M9G at 37°C. They were then diluted to $OD_{600}$ = 0.01 in M9G and grown at 37°C until $OD_{600}$ = 0.40, when RNA was extracted. For extraction, cells were spun for 30 seconds and the pellets were quickly resuspended in 1mL Trizol (Invitrogen, Life Technologies). We extracted the RNA using the manufacturer's protocol. The extracted RNA was next treated with Turbo DNase (Ambion, Life Technologies) and further purified using an RNA purification kit (Qiagen). Absence of DNA contamination was confirmed by PCR, where the lack of PCR product (about 100 bp in length) relative to a DNA-containing positive control was interpreted as evidence of DNA removal. The library preparations for RNA sequencing followed the ScriptSeq complete protocol (Illumina). Sequencing (50 bp single-end read) was performed on an Illumina HiSeq 2500. Transcript levels were mapped to the *E. coli* BW25113 genome in CLC Genomics Workbench 10 (mismatch cost=2; insertion cost=3, deletion cost=3, length fraction=0.8, similarity fraction=0.8). We note that data from samples collected on day one show particular systematic differences from samples collected on days two and three. To correct for the observed sample batch effects (associated with the day of RNA collection) when quantifying RNA-sequencing reads per gene across strains, we applied a generalized linear model in which dispersion was estimated using the Cox-Reid approach [93]. Sequencing reads for all primary deletions and corresponding *sup* strains have been deposited in the NCBI GEO database (accession number GSE85914).

**Data analysis of transcriptional changes**

To categorize genetic responses as restorative or compensatory adaptive evolution (R-AE and C-AE, respectively), we first used Venn diagrams to represent the R-AE responses. The transcription was studied through comparing changes (statistically significant with a fold change > 1.5) between the primary deletion and the WT strain, the primary deletion and an associated *sup* strain, and the WT strain and the same *sup* strain. In this set of comparisons, genes that possess statistically significant changes from the primary deletion (expression in the WT strain vs. the primary deletion strain) and from the adaptive evolution (expression in the primary deletion strain vs. the *sup* strain) only exhibit R-AE, while genes that possess statistically significant changes in all three cases, or the pairs of cases including the comparison of the expression in the WT vs. the *sup* strain as a member, exhibit C-AE (Fig 5B). For each primary deletion strain and its associated *sup* strains, we also compared the transcriptional state of the primary deletion and all *sup* strains with that of the WT strain (Fig 5C). We identified the set of genes that show a change between the primary deletion and the WT strain but not with any *sup* strain. All genes in this set exhibit R-AE, since the transcription rates of the adaptively evolved strains are in each case statistically indistinguishable from the WT strain. Meanwhile, genes demonstrating C-AE are contained in the set of genes that show differences between the WT expression level and that of all *sup* strains.

We used Gene Change Score (GCS) to more easily represent both R-AE and C-AE responses (as described in Fig 5). The GCS is a sum of Boolean variables $P + Q + R + S + T + U$, where each variable is 1 if true and 0 if false. The variables are defined as follows: $P$ indicates whether the gene expression change resulting from the gene deletion is statistically significant; $Q$ indicates whether the log fold change resulting from the gene deletion is larger than threshold; $R$ indicates whether the gene expression change resulting from the adaptive evolution is statistically significant; $S$ indicates whether the log fold change resulting from the adaptive evolution is larger than threshold; $T$ indicates whether the log fold changes have equal magnitude and opposite signs; and $U$ indicates whether the log fold changes have the same sign and the post-adaptation expression level is statistically significantly different from the WT expression level. Noting that only one of $T$ or $U$ can be true, the GCS is divided by 5 to normalize scores to be between zero and one. To highlight the difference between restorative and compensatory changes, the GCS is defined to be positive if the log fold changes are opposite in sign and negative otherwise.

To reduce sensitivity of the GCS to the specific threshold values, we calculated the scores from thresholds in a 5-by-5 grid of $\log_2$ fold change versus statistical significance ($p$-value). The grid is spaced evenly on the interval of the $95^{th}$–$99^{th}$ percentiles for both fold change and statistical significance (the percentiles are recalculated in for each initial deletion strain), and we averaged the scores over these 25 threshold combinations. Since C-AE and R-AE are mutually exclusive, we represented all scores on the same axis in Fig 6A by making the C-AE scores negative and R-AE scores positive with non-significant changes being scored as zero.

**Computational modeling**

**Quantifying transcriptional responses to deletions and *sup* mutations.** The RNA-sequencing data provided a snapshot of cellular transcription before and after the processes of gene deletion and adaptive evolution. In both cases, the transcriptional response was calculated by taking the difference between the final state (either deletion strain or evolved strain) and the initial state (WT strain or deletion strain, respectively) for each replicate, and then averaging over all replicates. The hypothetical transcriptional states—corresponding to implementing the *sup* mutations in primary deletion backgrounds different from the one in which they were selected (Fig 8)—were constructed by adding the average response to the measured transcriptional state of each primary deletion.

**Data used to map gene expression to growth rate.** We calculated growth rates from genome-wide gene expression levels using a data-driven, non-parametric mapping. Our central assumption in this mapping is that gene expression uniquely determines growth rate. This mapping was devised using the large collection of gene expression data with associated growth rate measurements for *E. coli* K12 in [59], which includes 589 genome-wide measurements across a variety of conditions. We added our 57 RNA-sequencing measurements (19 different conditions with three replicates each) resulting in a total of $m$ = 646 gene expression experiments with associated growth rate. To accurately measure transcriptional correlations between genes, along with these 646 datasets, we included 1,607 additional gene expression profiles cataloged in [59] for which no growth rate is available, for a total of $N$ = 2,253 expression profiles. We note that we included growth rate data from a variety of conditions beyond defined media for two reasons: The first was to incorporate as many growth rate measurements as possible since growth measurements with associated growth rates are relatively rare. The second was to help resolve gene-gene correlations. As we will show, it is these correlations that allow us to represent the high-dimensional transcriptomic data in a low-dimensional representation.

**Estimation of eigengenes.** To quantitatively characterize the correlations between genes, we constructed *eigengenes* [94]. First, we calculated all pairwise (Pearson) correlations between genes using the $N$ = 2,253 expression profiles, which were quantile-normalized [95] to make the datasets comparable. The number of genes common to all expression profiles in our data is $G$ = 3,969, resulting in a 3,969 x 3,969 correlation matrix, which has $N$ nonzero eigenvalues. We decomposed this matrix using singular value decomposition to calculate the eigenvalues and eigenvectors. Each nonzero eigenvalue quantifies the amount of variance along its associated eigenvector, which together reduce the dimension of the space from $G$ to $N$. The eigenvectors are themselves linear combinations of genes that define how related one gene's expression is to another across the database. We identify the eigenvectors obtained through this process with the concept of eigengenes put forward in [94].

**Growth rate prediction from eigengenes.** Using the $m$ = 646 expression profiles with associated growth rate, we performed $k$-nearest-neighbors (KNN) regression [60, 61]. Figure 7A illustrates the idea behind KNN regression: each gene expression profile is expressed in terms of eigengenes by projecting the corresponding vector of gene expression onto the eigengenes derived above. To predict an unknown growth rate ($\widehat{f_0}$) associated with an expression profile ($\boldsymbol{g_0}$ **=** $(g^{(1)}, g^{(2)}, \ldots, g^{(N)})$), we calculated the (Euclidean) distance-weighted average of the growth rates associated with the most similar gene expression profiles in our dataset. We define $k$ to be a fixed number of neighbors ($k$ = 8 in our simulations), $\mathbb{b}$ to denote the set of $k$ expression profiles with the smallest distance to $\boldsymbol{g_0}$, $i$ and $j$ to be indices over $\mathbb{b}$, and $\boldsymbol{g_i}$ and $f_i$ to be the known $i^{th}$-neighbor gene expression and growth rate, respectively. The weight, $w_i$, associated with the distance to the $i^{th}$ neighbor, is

$$w_i = \frac{\|\boldsymbol{g}_i - \boldsymbol{g_0}\|^{-2}}{\sum_{j\in\mathbb{b}}\|\boldsymbol{g}_j - \boldsymbol{g_0}\|^{-2}}. \qquad (1)$$

Using these $w_i$, the growth rate $\widehat{f_0}$ was estimated as $\widehat{f_0} = \sum_{i\in\mathbb{b}} w_i\, f_i$. We note that our results are not sensitive to the number of neighbors $k$ in our model.

**Prediction criterion for model selection.** The estimated growth rate accuracy was assessed through stratified five-fold cross-validation, in which we divided the data into representative fifths based on growth rate (called folds). Four folds of the data were then used to predict the growth rate of the remaining fold. The prediction was performed with each fold left out so that every experiment had a measured and predicted growth rate. The residual error was scored by the sum of squared errors between the predicted and actual values of the growth rate:

$$r = \sum_{i=1}^{m} \left\| \hat{f}_i - f_i \right\|^2, \qquad (2)$$

where *i* is an index over the set of all *m* = 646 experiments. Accurate models are characterized by small values of *r*, providing a measure for comparing models.

**Accounting for noise in gene expression.** Eigengenes associated with small eigenvalues exhibit large sensitivity to noise, whereas those associated with large eigenvalues exhibit almost none. We accounted for this by simulating noisy data based on $\sigma^2 = \mu(1 + \beta) + \alpha\mu^2$, where $\sigma$ is the measured variance in the expression of a gene, $\mu$ is the measured mean expression of a gene, and $\alpha$ and $\beta$ are hyper-parameters estimated through the DESeq software [96] ($\alpha = 0.217 \pm 0.004$ and $\beta = 0.539 \pm 0.040$). Taking the sample mean value for each gene to be an estimate of $\mu$, we calculated $\sigma$ and simulated pseudo-measurements drawn from a Gaussian distribution. The pseudo-measurements were created for each gene in the profile and projected onto the eigengenes resulting in pseudo-profiles. A total of 4*m* such pseudo-profiles were included in the cross-validation scheme described above.

**Addition of zero-growth pseudo-counts.** While naive KNN regression performs better when interpolating within the range of data than when extrapolating beyond it, properly assigning zero-growth rate to non-physical expression states (negative or unrealistically large expression) improves extrapolation. Specifically, we augmented our training set with zero-growth non-physical expression states (called pseudo-counts). These pseudo-counts were chosen from a spherical shell around the center of mass $\bar{\boldsymbol{g}}$ of the observed expression levels,

$$\bar{\boldsymbol{g}} = \frac{1}{m}\sum_{j=1}^{m} \boldsymbol{g}_j, \qquad (3)$$

where the radius of the shell was taken to be $d = \max_{1 \le i,j \le m} \left\| \boldsymbol{g}_i - \boldsymbol{g}_j \right\|^2$. To generate the gene expression of the pseudo-counts, we randomly selected *m*(*m*–1)/16 distinct pairs of measurements $\boldsymbol{g}_i$ and $\boldsymbol{g}_j$, and took as pseudo-counts the points on the spherical shell at the intersection with the line connecting them.

**Selection of eigengenes that best predict growth.** We note that the distribution of eigenvalues spans many orders of magnitude, with most of the variance being captured by a small minority of eigengenes. This observation suggests that growth rate can be predicted both accurately and robustly using only a few eigengenes. We use the term *n*-eigengene model to refer to a KNN model in which the predictions are based on a fixed set of *n* eigengenes. Finding the best set of eigengenes would in principle require the testing of all $\sum_{n=1}^{N} N!/[(N-n)!\,(n)!]$ combinations of eigengenes, which is computationally impractical. To circumvent this difficulty, starting with all *n*-eigengene models for *n* = 1, we adopted a forward selection heuristic [97] in which: 1) we cross-validated the constructed set of *n*-eigengene models and selected the best; 2) we constructed all ($n$+1)-eigengene models formed by pairing the best $n$-eigengene model with each of the remaining eigengenes; 3) increasing $n$ by 1 at each iteration, we repeated 1) and 2) until the accuracy stops improving sufficiently to justify the addition of another eigengene. As more eigengenes are added in the procedure above, prediction of the growth rate improves in tandem with the risk of overfitting. The accuracy was balanced against overfitting by comparing the marginal improvement of the best $n$-eigengene model over the best ($n$–1)-eigengene model with the number of unoccupied bins in the gene expression space. Specifically, the gene expression profiles were divided evenly into $N_f$ = 13 bins according to their growth rates, and then further divided into decile bins according to the projection along each of the $n$ eigengenes in the model (unless this resulted in neighboring deciles separated by less than 10% of their mean projection, in which case the bins were merged). We define $N_i$ to be the number of bins for the $i^{th}$ eigengene in the model and $N_o$ to be the number bins (out of the total of up to $10^n$ x $N_f$ bins) that are occupied by at least one of the $m$ expression profiles. To find the smallest set of eigengenes that effectively predict growth rate, we minimized

$$\underset{n}{\operatorname{argmin}}\, r(n) - \lambda L(n), \tag{4}$$

where $r$ is as in Eq (2), $L = 2\ln N_o - \ln N_f - \sum_{i=1}^{n+1} \ln N_i$ is the log-occupation ratio, and $\lambda$ is a regularization parameter set to be 0.1 (and we tested that the results are not sensitive to this choice). For our data, this minimization resulted in $n$ = 9 eigengenes, which is a significant reduction from the total of 2,253 eigengenes associated with the 3,969 common genes in our database.

**Out-of-sample cross-validation of growth rate as a function of the training data.** In the validation presented in Fig 7C*,* the 2,198 experiments curated in [59] were augmented with data systematically gathered from the Gene Expression Omnibus (GEO) database for keywords “growth rate” and “Escherichia coli” (access date: Nov. 14, 2017). The selected experiments are cataloged in S4 Table, where [59] indicates the data used to build the model in that reference. We selected experiments that measured absolute transcription levels and growth rate, including recent RNA-Seq experiments [64], but excluded two-channel experiments and experiments conducted before 2014 (as they would have been included in [59] dataset). After gathering the data, growth rates were obtained from the associated publications, either by interpolating growth rates from the figures or taking them directly from a table in the paper materials. If the processed gene expression data was reported in the paper we used that, otherwise we downloaded the data from the GEO series page.

The datasets were then filtered according to the number of genes they shared with our starting dataset ([59]). The GEO series were ordered by the shared number of genes shared and selected the series that shared the greatest number of genes. Then, the set of shared genes was updated. The process of ordering series by the number of shared genes, selecting the series that shared the most genes, and updating the shared gene set was repeated until all series were included. To contextualize the accuracy of the KNN analysis, the four datasets described in S4 Table were subjected to the KNN analysis. The prediction accuracy of our growth rate data was quantified by the coefficient of determination ($R^2$) as well as the number of experiments predicted within 25% and 10% of their measured growth rate. In addition to the out-of-sample prediction, we selected the replicates of one strain as a test set out, used the remaining data to train the KNN model, and again measured the number of experiments predicted within 25% and 10% of their measured growth rate. We note that including all data significantly diminishes the number of shared genes from 4,189 to 1,683, as indicated in S4 Table. For presentation in the main text, we selected the largest number of experiments that shared more than half of the genes in *E. coli* (Dataset D2 in S4 Table).

Funding Information
The funders had no role in study design, data collection and analysis, decision to publish, or preparation of the manuscript.

**References**

1. Shuman HA, Silhavy TJ. The art and design of genetic screens: *Escherichia coli*. Nat Rev Genet. 2003;4(6): 419-31. doi: 10.1038/nrg1087.
2. Fong SS, Palsson BO. Metabolic gene-deletion strains of *Escherichia coli* evolve to computationally predicted growth phenotypes. Nat Genet. 2004;36(10): 1056-8. doi: 10.1038/ng1432.
3. Barrick JE, Lenski RE. Genome dynamics during experimental evolution. Nat Rev Genet. 2013;14(12): 827-39. doi: 10.1038/nrg3564.
4. Herring CD, Raghunathan A, Honisch C, Patel T, Applebee MK, Joyce AR, et al. Comparative genome sequencing of *Escherichia coli* allows observation of bacterial evolution on a laboratory timescale. Nat Genet. 2006;38(12):1406-12. Epub 2006/11/07. doi: 10.1038/ng1906.
5. Conrad TM, Frazier M, Joyce AR, Cho BK, Knight EM, Lewis NE, et al. RNA polymerase mutants found through adaptive evolution reprogram *Escherichia coli* for optimal growth in minimal media. Proc Natl Acad Sci U S A. 2010;107(47): 20500-5. doi: 10.1073/pnas.0911253107.
6. LaCroix RA, Sandberg TE, O'Brien EJ, Utrilla J, Ebrahim A, Guzman GI, et al. Use of adaptive laboratory evolution to discover key mutations enabling rapid growth of *Escherichia coli* K-12 MG1655 on glucose minimal medium. Appl Environ Microbiol. 2015;81(1): 17-30. doi: 10.1128/AEM.02246-14.
7. Kishimoto T, Iijima L, Tatsumi M, Ono N, Oyake A, Hashimoto T, et al. Transition from positive to neutral in mutation fixation along with continuing rising fitness in thermal adaptive evolution. PLoS Genet. 2010;6(10):e1001164. Epub 2010/10/27. doi: 10.1371/journal.pgen.1001164.
8. Minty JJ, Lesnefsky AA, Lin F, Chen Y, Zaroff TA, Veloso AB, et al. Evolution combined with genomic study elucidates genetic bases of isobutanol tolerance in *Escherichia coli*. Microb Cell Fact. 2011;10:18. Epub 2011/03/26. doi: 10.1186/1475-2859-10-18.
9. Wang L, Spira B, Zhou Z, Feng L, Maharjan RP, Li X, et al. Divergence involving global regulatory gene mutations in an *Escherichia coli* population evolving under phosphate limitation. Genome Biol Evol. 2010;2:478-87. Epub 2010/07/20. doi: 10.1093/gbe/evq035.
10. Lee DH, Palsson BO. Adaptive evolution of Escherichia coli K-12 MG1655 during growth on a nonnative carbon source, L-1,2-propanediol. Appl Environ Microbiol. 2010;76(13):4158-68. Epub 2010/05/04. doi: 10.1128/AEM.00373-10.
11. Toprak E, Veres A, Michel JB, Chait R, Hartl DL, Kishony R. Evolutionary paths to antibiotic resistance under dynamically sustained drug selection. Nat Genet. 2011;44(1):101-5. Epub 2011/12/20. doi: 10.1038/ng.1034.
12. Sandberg TE, Long CP, Gonzalez JE, Feist AM, Antoniewicz MR, Palsson BO. Evolution of *E. coli* on [U-13C]glucose reveals a negligible isotopic influence on metabolism and physiology. PLoS One. 2016;11(3):e0151130. Epub 2016/03/11. doi: 10.1371/journal.pone.0151130.

13. He A, Penix SR, Basting PJ, Griffith JM, Creamer KE, Camperchioli D, et al. Acid evolution of *Escherichia coli* K-12 eliminates amino acid decarboxylases and reregulates catabolism. Appl Environ Microbiol. 2017;83(12). doi: 10.1128/AEM.00442-17.
14. Tenaillon O, Rodriguez-Verdugo A, Gaut RL, McDonald P, Bennett AF, Long AD, et al. The molecular diversity of adaptive convergence. Science. 2012;335(6067):457-61. doi: 10.1126/science.1212986.
15. Auriol C, Bestel-Corre G, Claude JB, Soucaille P, Meynial-Salles I. Stress-induced evolution of *Escherichia coli* points to original concepts in respiratory cofactor selectivity. Proc Natl Acad Sci U S A. 2011;108(4):1278-83. Epub 2011/01/06. doi: 10.1073/pnas.1010431108.
16. Chou HH, Chiu HC, Delaney NF, Segre D, Marx CJ. Diminishing returns epistasis among beneficial mutations decelerates adaptation. Science. 2011;332(6034):1190-2. Epub 2011/06/04. doi: 10.1126/science.1203799.
17. Charusanti P, Conrad TM, Knight EM, Venkataraman K, Fong NL, Xie B, et al. Genetic basis of growth adaptation of *Escherichia coli* after deletion of *pgi*, a major metabolic gene. PLoS Genet. 2010;6(11): e1001186. doi: 10.1371/journal.pgen.1001186.
18. Blank D, Wolf L, Ackermann M, Silander OK. The predictability of molecular evolution during functional innovation. Proc Natl Acad Sci U S A. 2014;111(8):3044-9. doi: 10.1073/pnas.1318797111.
19. Motter AE. Improved network performance via antagonism: From synthetic rescues to multi-drug combinations. Bioessays. 2010;32(3): 236-45. doi: 10.1002/bies.200900128.
20. Chatterjee R, Millard CS, Champion K, Clark DP, Donnelly MI. Mutation of the *ptsG* gene results in increased production of succinate in fermentation of glucose by *Escherichia coli*. Appl Environ Microbiol. 2001;67(1): 148-54. doi: 10.1128/AEM.67.1.148-154.2001.
21. Gupta S, Clark DP. *Escherichia coli* derivatives lacking both alcohol dehydrogenase and phosphotransacetylase grow anaerobically by lactate fermentation. J Bacteriol. 1989;171(7): 3650-5.
22. Motter AE, Gulbahce N, Almaas E, Barabasi AL. Predicting synthetic rescues in metabolic networks. Mol Syst Biol. 2008;4: 168. doi: 10.1038/msb.2008.1.
23. Trinh CT, Carlson R, Wlaschin A, Srienc F. Design, construction and performance of the most efficient biomass producing *E. coli* bacterium. Metab Eng. 2006;8(6): 628-38. doi: 10.1016/j.ymben.2006.07.006.
24. Fischer E, Sauer U. Large-scale in vivo flux analysis shows rigidity and suboptimal performance of *Bacillus subtilis* metabolism. Nat Genet. 2005;37(6): 636-40. doi: 10.1038/ng1555.
25. Reguly T, Breitkreutz A, Boucher L, Breitkreutz BJ, Hon GC, Myers CL, et al. Comprehensive curation and analysis of global interaction networks in *Saccharomyces cerevisiae*. J Biol. 2006;5(4): 11. doi: 10.1186/jbiol36.
26. Cornelius SP, Lee JS, Motter AE. Dispensability of *Escherichia coli*'s latent pathways. Proc Natl Acad Sci U S A. 2011;108(8): 3124-9. doi: 10.1073/pnas.1009772108.
27. Butland G, Babu M, Diaz-Mejia JJ, Bohdana F, Phanse S, Gold B, et al. eSGA: *E. coli* synthetic genetic array analysis. Nat Methods. 2008;5(9): 789-95. doi: 10.1038/nmeth.1239.
28. Typas A, Nichols RJ, Siegele DA, Shales M, Collins SR, Lim B, et al. High-throughput, quantitative analyses of genetic interactions in *E. coli*. Nat Methods. 2008;5(9): 781-7.
29. Woods R, Schneider D, Winkworth CL, Riley MA, Lenski RE. Tests of parallel molecular evolution in a long-term experiment with *Escherichia coli*. Proc Natl Acad Sci U S A. 2006;103(24): 9107-12. doi: 10.1073/pnas.0602917103.

30. Deatherage DE, Barrick JE. Identification of mutations in laboratory-evolved microbes from next-generation sequencing data using breseq. Methods Mol Biol. 2014;1151:165-88. doi: 10.1007/978-1-4939-0554-6_12.
31. Opalka N, Brown J, Lane WJ, Twist KA, Landick R, Asturias FJ, et al. Complete structural model of *Escherichia coli* RNA polymerase from a hybrid approach. PLoS Biol. 2010;8(9): e1000483. doi: 10.1371/journal.pbio.1000483.
32. Campbell EA, Korzheva N, Mustaev A, Murakami K, Nair S, Goldfarb A, et al. Structural mechanism for rifampicin inhibition of bacterial RNA polymerase. Cell. 2001;104(6): 901-12.
33. Ederth J, Mooney RA, Isaksson LA, Landick R. Functional interplay between the jaw domain of bacterial RNA polymerase and allele-specific residues in the product RNA-binding pocket. J Mol Biol. 2006;356(5): 1163-79. doi: 10.1016/j.jmb.2005.11.080.
34. Goldstein BP. Resistance to rifampicin: a review. J Antibiot (Tokyo). 2014;67(9): 625-30. doi: 10.1038/ja.2014.107.
35. Jin DJ, Gross CA. Mapping and sequencing of mutations in the *Escherichia coli rpoB* gene that lead to rifampicin resistance. J Mol Biol. 1988;202(1): 45-58.
36. Zhou YN, Lubkowska L, Hui M, Court C, Chen S, Court DL, et al. Isolation and characterization of RNA polymerase *rpoB* mutations that alter transcription slippage during elongation in *Escherichia coli*. J Biol Chem. 2013;288(4): 2700-10. doi: 10.1074/jbc.M112.429464.
37. Chib S, Ali F, Seshasayee ASN. Genomewide mutational diversity in *Escherichia coli* population evolving in prolonged stationary phase. mSphere. 2017;2(3): e00059-17. Epub 2017/06/02. doi: 10.1128/mSphere.00059-17.
38. Feng Y, Degen D, Wang X, Gigliotti M, Liu S, Zhang Y, et al. Structural basis of transcription inhibition by CBR hydroxamidines and CBR pyrazoles. Structure. 2015;23(8): 1470-81. doi: 10.1016/j.str.2015.06.009.
39. Malinen AM, Nandymazumdar M, Turtola M, Malmi H, Grocholski T, Artsimovitch I, et al. CBR antimicrobials alter coupling between the bridge helix and the beta subunit in RNA polymerase. Nat Commun. 2014;5: 3408. doi: 10.1038/ncomms4408.
40. Weilbaecher R, Hebron C, Feng G, Landick R. Termination-altering amino acid substitutions in the beta' subunit of *Escherichia coli* RNA polymerase identify regions involved in RNA chain elongation. Genes Dev. 1994;8(23): 2913-27.
41. Ross W, Vrentas CE, Sanchez-Vazquez P, Gaal T, Gourse RL. The magic spot: a ppGpp binding site on *E. coli* RNA polymerase responsible for regulation of transcription initiation. Mol Cell. 2013;50(3):420-9. doi: 10.1016/j.molcel.2013.03.021.
42. Ha KS, Toulokhonov I, Vassylyev DG, Landick R. The NusA N-terminal domain is necessary and sufficient for enhancement of transcriptional pausing via interaction with the RNA exit channel of RNA polymerase. J Mol Biol. 2010;401(5): 708-25. doi: 10.1016/j.jmb.2010.06.036.
43. Fraenkel DG. Glycolysis, pentose phosphate pathway, and Entner–Doudoroff pathway. In: Neidhardt FC, editor. *Escherichia coli* and *Salmonella typhimurium*: Cellular and Molecular Biology. 1. Washington, D.C.: ASM Press; 1987. p. 142-50.
44. Fuhrman LK, Wanken A, Nickerson KW, Conway T. Rapid accumulation of intracellular 2-keto-3-deoxy-6-phosphogluconate in an Entner-Doudoroff aldolase mutant results in bacteriostasis. FEMS Microbiol Lett. 1998;159(2): 261-6.
45. Nishikawa T, Gulbahce N, Motter AE. Spontaneous reaction silencing in metabolic optimization. PLoS Comput Biol. 2008;4(12): e1000236. doi: 10.1371/journal.pcbi.1000236.

46. Mi H, Huang X, Muruganujan A, Tang H, Mills C, Kang D, et al. PANTHER version 11: expanded annotation data from Gene Ontology and Reactome pathways, and data analysis tool enhancements. Nucleic Acids Res. 2017;45(D1): D183-D9. doi: 10.1093/nar/gkw1138.
47. Hor L, Dobson RC, Downton MT, Wagner J, Hutton CA, Perugini MA. Dimerization of bacterial diaminopimelate epimerase is essential for catalysis. J Biol Chem. 2013;288(13): 9238-48. doi: 10.1074/jbc.M113.450148.
48. Kwon O, Hudspeth ME, Meganathan R. Anaerobic biosynthesis of enterobactin *Escherichia coli*: regulation of *entC* gene expression and evidence against its involvement in menaquinone (vitamin K2) biosynthesis. J Bacteriol. 1996;178(11): 3252-9.
49. Gama-Castro S, Salgado H, Santos-Zavaleta A, Ledezma-Tejeida D, Muniz-Rascado L, Garcia-Sotelo JS, et al. RegulonDB version 9.0: high-level integration of gene regulation, coexpression, motif clustering and beyond. Nucleic Acids Res. 2016;44(D1): D133-D43. doi: 10.1093/nar/gkv1156.
50. Adams MD, Wagner LM, Graddis TJ, Landick R, Antonucci TK, Gibson AL, et al. Nucleotide sequence and genetic characterization reveal six essential genes for the LIV-I and LS transport systems of *Escherichia coli*. J Biol Chem. 1990;265(20): 11436-43.
51. Beisel CL, Storz G. The base-pairing RNA spot 42 participates in a multioutput feedforward loop to help enact catabolite repression in *Escherichia coli*. Mol Cell. 2011;41(3): 286-97. doi: 10.1016/j.molcel.2010.12.027.
52. Utrilla J, O'Brien EJ, Chen K, McCloskey D, Cheung J, Wang H, et al. Global rebalancing of cellular resources by pleiotropic point mutations illustrates a multi-scale mechanism of adaptive evolution. Cell Syst. 2016;2(4): 260-71. doi: 10.1016/j.cels.2016.04.003.
53. Ma Z, Gong S, Richard H, Tucker DL, Conway T, Foster JW. GadE (YhiE) activates glutamate decarboxylase-dependent acid resistance in *Escherichia coli* K-12. Mol Microbiol. 2003;49(5): 1309-20.
54. Masuda N, Church GM. Regulatory network of acid resistance genes in *Escherichia coli*. Mol Microbiol. 2003;48(3): 699-712.
55. Tucker DL, Tucker N, Ma Z, Foster JW, Miranda RL, Cohen PS, et al. Genes of the GadX-GadW regulon in *Escherichia coli*. J Bacteriol. 2003;185(10): 3190-201.
56. Ma Z, Richard H, Tucker DL, Conway T, Foster JW. Collaborative regulation of *Escherichia coli* glutamate-dependent acid resistance by two AraC-like regulators, GadX and GadW (YhiW). J Bacteriol. 2002;184(24): 7001-12.
57. Yoshida T, Ueguchi C, Mizuno T. Physical map location of a set of *Escherichia coli* genes (*hde*) whose expression is affected by the nucleoid protein H-NS. J Bacteriol. 1993;175(23): 7747-8.
58. Weber H, Polen T, Heuveling J, Wendisch VF, Hengge R. Genome-wide analysis of the general stress response network in *Escherichia coli*: sigmaS-dependent genes, promoters, and sigma factor selectivity. J Bacteriol. 2005;187(5):1591-603. doi: 10.1128/JB.187.5.1591-1603.2005.
59. Carrera J, Estrela R, Luo J, Rai N, Tsoukalas A, Tagkopoulos I. An integrative, multi-scale, genome-wide model reveals the phenotypic landscape of *Escherichia coli*. Mol Syst Biol. 2014;10(7): 735. doi: 10.15252/msb.20145108.
60. Altman NS. An introduction to kernel and nearest-neighbor nonparametric regression. Am Stat. 1992;46: 175-85.
61. Pedragosa F, Varoquaux G, Gramfort A, Michel V, Thirion B, Grisel O, et al. Scikit-learn: Machine Learning in Python. J Mach Learn Res. 2011;12: 2825-30.
62. Stern DL. The genetic causes of convergent evolution. Nat Rev Genet. 2013;14(11):751-64. doi: 10.1038/nrg3483.

63. Conrad TM, Lewis NE, Palsson BO. Microbial laboratory evolution in the era of genome-scale science. Mol Syst Biol. 2011;7(1): 509. doi: 10.1038/msb.2011.42.
64. Kram KE, Geiger C, Ismail WM, Lee H, Tang H, Foster PL, et al. Adaptation of *Escherichia coli* to long-term serial passage in complex medium: evidence of parallel evolution. mSystems. 2017;2(2): e00192-16. doi: 10.1128/mSystems.00192-16.
65. Stoebel DM, Dorman CJ. The effect of mobile element IS10 on experimental regulatory evolution in *Escherichia coli*. Mol Biol Evol. 2010;27(9): 2105-12. doi: 10.1093/molbev/msq101.
66. Ho MX, Hudson BP, Das K, Arnold E, Ebright RH. Structures of RNA polymerase-antibiotic complexes. Curr Opin Struct Biol. 2009;19(6): 715-23. doi: 10.1016/j.sbi.2009.10.010.
67. Zhang Y, Feng Y, Chatterjee S, Tuske S, Ho MX, Arnold E, et al. Structural basis of transcription initiation. Science. 2012;338(6110): 1076-80. doi: 10.1126/science.1227786.
68. Goodarzi H, Hottes AK, Tavazoie S. Global discovery of adaptive mutations. Nat Methods. 2009;6(8): 581-3. doi: 10.1038/nmeth.1352.
69. Barrick JE, Yu DS, Yoon SH, Jeong H, Oh TK, Schneider D, et al. Genome evolution and adaptation in a long-term experiment with *Escherichia coli*. Nature. 2009;461(7268): 1243-7. doi: 10.1038/nature08480.
70. Schneider D, Duperchy E, Coursange E, Lenski RE, Blot M. Long-term experimental evolution in *Escherichia coli*. IX. Characterization of insertion sequence-mediated mutations and rearrangements. Genetics. 2000;156(2): 477-88.
71. Liu M, Durfee T, Cabrera JE, Zhao K, Jin DJ, Blattner FR. Global transcriptional programs reveal a carbon source foraging strategy by *Escherichia coli*. J Biol Chem. 2005;280(16): 15921-7. doi: 10.1074/jbc.M414050200.
72. Giangrossi M, Zattoni S, Tramonti A, De Biase D, Falconi M. Antagonistic role of H-NS and GadX in the regulation of the glutamate decarboxylase-dependent acid resistance system in *Escherichia coli*. J Biol Chem. 2005;280(22): 21498-505. doi: 10.1074/jbc.M413255200.
73. Hommais F, Krin E, Laurent-Winter C, Soutourina O, Malpertuy A, Le Caer JP, et al. Large-scale monitoring of pleiotropic regulation of gene expression by the prokaryotic nucleoid-associated protein, H-NS. Mol Microbiol. 2001;40(1): 20-36.
74. Grainger DC. Structure and function of bacterial H-NS protein. Biochem Soc Trans. 2016;44(6): 1561-9. doi: 10.1042/BST20160190.
75. Seo SW, Kim D, O'Brien EJ, Szubin R, Palsson BO. Decoding genome-wide GadEWX-transcriptional regulatory networks reveals multifaceted cellular responses to acid stress in *Escherichia coli*. Nat Commun. 2015;6:7970. doi: 10.1038/ncomms8970.
76. Typas A, Becker G, Hengge R. The molecular basis of selective promoter activation by the sigmaS subunit of RNA polymerase. Mol Microbiol. 2007;63(5): 1296-306. doi: 10.1111/j.1365-2958.2007.05601.x.
77. Ko M, Park C. H-NS-Dependent regulation of flagellar synthesis is mediated by a LysR family protein. J Bacteriol. 2000;182(16): 4670-2.
78. Li H, Sourjik V. Assembly and stability of flagellar motor in *Escherichia coli*. Mol Microbiol. 2011;80(4): 886-99. doi: 10.1111/j.1365-2958.2011.07557.x.
79. Soutourina O, Kolb A, Krin E, Laurent-Winter C, Rimsky S, Danchin A, et al. Multiple control of flagellum biosynthesis in *Escherichia coli*: role of H-NS protein and the cyclic AMP-catabolite activator protein complex in transcription of the *flhDC* master operon. J Bacteriol. 1999;181(24): 7500-8.
80. Typas A, Sourjik V. Bacterial protein networks: properties and functions. Nat Rev Microbiol. 2015;13(9): 559-72. doi: 10.1038/nrmicro3508.

81. Dong T, Schellhorn HE. Control of RpoS in global gene expression of *Escherichia coli* in minimal media. Mol Genet Genomics. 2009;281(1):19-33. doi: 10.1007/s00438-008-0389-3.
82. Makinoshima H, Aizawa S, Hayashi H, Miki T, Nishimura A, Ishihama A. Growth phase-coupled alterations in cell structure and function of *Escherichia coli*. J Bacteriol. 2003;185(4):1338-45.
83. Dong T, Yu R, Schellhorn H. Antagonistic regulation of motility and transcriptome expression by RpoN and RpoS in *Escherichia coli*. Mol Microbiol. 2011;79(2):375-86. doi: 10.1111/j.1365-2958.2010.07449.x.
84. Szamecz B, Boross G, Kalapis D, Kovacs K, Fekete G, Farkas Z, et al. The genomic landscape of compensatory evolution. PLoS Biol. 2014;12(8):e1001935. Epub 2014/08/27. doi: 10.1371/journal.pbio.1001935.
85. Costanzo M, VanderSluis B, Koch EN, Baryshnikova A, Pons C, Tan G, et al. A global genetic interaction network maps a wiring diagram of cellular function. Science. 2016;353(6306): aaf1420. doi: 10.1126/science.aaf1420.
86. van Leeuwen J, Pons C, Mellor JC, Yamaguchi TN, Friesen H, Koschwanez J, et al. Exploring genetic suppression interactions on a global scale. Science. 2016;354(6312): aag0839. doi: 10.1126/science.aag0839.
87. Partow S, Hyland PB, Mahadevan R. Synthetic rescue couples NADPH generation to metabolite overproduction in *Saccharomyces cerevisiae*. Metab Eng. 2017 Sep; 43(Pt A):64–70. doi: 10.1016/j.ymben.2017.08.004.
88. Chait R, Craney A, Kishony R. Antibiotic interactions that select against resistance. Nature. 2007;446(7136): 668-71. doi: 10.1038/nature05685.
89. Baba T, Ara T, Hasegawa M, Takai Y, Okumura Y, Baba M, et al. Construction of *Escherichia coli* K-12 in-frame, single-gene knockout mutants: the Keio collection. Mol Syst Biol. 2006;2(1): 2006.0008. doi: 10.1038/msb4100050.
90. Link AJ, Phillips D, Church GM. Methods for generating precise deletions and insertions in the genome of wild-type *Escherichia coli*: application to open reading frame characterization. J Bacteriol. 1997;179(20):6228-37.
91. Grenier F, Matteau D, Baby V, Rodrigue S. Complete genome sequence of *Escherichia coli* BW25113. Genome Announc. 2014;2(5): e01038-14. doi: 10.1128/genomeA.01038-14.
92. Kearse M, Moir R, Wilson A, Stones-Havas S, Cheung M, Sturrock S, et al. Geneious Basic: an integrated and extendable desktop software platform for the organization and analysis of sequence data. Bioinformatics. 2012;28(12): 1647-9. doi: 10.1093/bioinformatics/bts199.
93. Robinson MD, McCarthy DJ, Smyth GK. edgeR: a Bioconductor package for differential expression analysis of digital gene expression data. Bioinformatics. 2010;26(1): 139-40. doi: 10.1093/bioinformatics/btp616.
94. Alter O, Brown PO, Botstein D. Singular value decomposition for genome-wide expression data processing and modeling. Proc Natl Acad Sci U S A. 2000;97(18): 10101-6.
95. Bolstad BM, Irizarry RA, Astrand M, Speed TP. A comparison of normalization methods for high density oligonucleotide array data based on variance and bias. Bioinformatics. 2003;19(2):185-93.
96. Love MI, Huber W, Anders S. Moderated estimation of fold change and dispersion for RNA-seq data with DESeq2. Genome Biol. 2014;15(12): 550. doi: 10.1186/s13059-014-0550-8.
97. Efron B, Hastie T, Johnstone I, Tibshirani R. Least angle regression. Ann Stat. 2004;32: 407-99.
98. McClure R, Balasubramanian D, Sun Y, Bobrovskyy M, Sumby P, Genco CA, et al. Computational analysis of bacterial RNA-Seq data. Nucleic Acids Res. 2013;41(14):e140. doi: 10.1093/nar/gkt444.

## Tables and Figures

**Table 1. Mapped genetic changes in 12 *Fast* strains evolved from wild type and 16 *sup* strains evolved from mutant backgrounds.**

| Strain | Reference Sequence Location | Mutation | Coding / non-coding change(s) [a] | Locus ID |
|---|---|---|---|---|
| **Strains evolved from WT parent** | | | | |
| Fast_1a | 4,176,719 | G > A | *rpoC*(R481H) | BW25113_3988 |
| Fast_1b | 1,749,977 | G > A | *pykF*(C8Y) | BW25113_1676 |
| Fast_2a | 1,750,065 | ΔT | *pykF*(H37fs) | BW25113_1676 |
| Fast_2b | 1,750,065 | ΔT | *pykF*(H37fs) | BW25113_1676 |
| | 608,104 | G > T | *fepA*(-154) / *fes*(-89) | BW25113_0584-5 |
| Fast_3a & 3b | 1,759,946 | IS*5*(+) | *pykF*(-3::IS*5*) | BW25113_1676 |
| Fast_4a & 4b | 4,174,281 | A > C | *rpoB*(T1037P) | BW25113_3987 |
| Fast_5a | 1,750,309 | IS*5*(+) | *pykF*(V119::IS*5*) | BW25113_1676 |
| | 647,643 | G > T | *citC*(-218) / *citA*(-161) | BW25113_0618-9 |
| Fast_5b | 1,750,309 | IS*5*(+) | *pykF*(V119::IS*5*) | BW25113_1676 |
| Fast_6a | 4,178,881 | G > A | *rpoC*(E1202K) | BW25113_3988 |
| Fast_6b | 4,178,881 | G > A | *rpoC*(E1202K) | BW25113_3988 |
| | 2,471,573 | C > T | *dsdX*(S83F) | BW25113_2365 |
| **Strains evolved from mutant parents** | | | | |
| Δ*zwf sup*1 * [b] | 4,178,209 | C > A | *rpoC*(R978S) | BW25113_3988 |
| | 1,301,660 | IS*5*(+) | *clsA(*L414*::*IS*5)* | BW25113_1249 |
| Δ*zwf sup*2 | 4,171,264 | A > G | *rpoB*(Q31R) | BW25113_3987 |
| | 766,917 | IS*5*(+) | *cydA(*L2::IS*5)* | BW25113_0733 |
| Δ*zwf sup*3 | 1,107,358 | C > T | *opgH*(A347V) | BW25113_1049 |
| | 4,172,719 | A > G | *rpoB*(D516G) | BW25113_3987 |
| Δ*zwf sup*4 | 155,376 | C > T | *pcnB*(D80N) | BW25113_0143 |
| | 4,172,719 | A > G | *rpoB*(D516G) | BW25113_3987 |
| Δ*ppk sup*1 | 4,178,896 | $(TAGAACGTG)_{1\rightarrow 2}$ | *rpoC*(G1207VERG) | BW25113_3988 |
| Δ*ppk sup*2 | 4,178,210 | G > C | *rpoC*(R978P) | BW25113_3988 |
| Δ*ppk sup*3 | 4,173,134 | C > G | *rpoB*(D654E) | BW25113_3987 |
| Δ*dapF sup*1 | 4,174,048 | A > G | *rpoB*(D959G) | BW25113_3987 |
| | 928,451 | C > A | *lrp*(T134N) | BW25113_0889 |
| Δ*dapF sup*2 | 4,173,309 | G > T | *rpoB*(G713C) | BW25113_3987 |
| | 928,451 | C > A | *lrp*(T134N) | BW25113_0889 |
| Δ*dapF sup*3 | 3,809,168 | ΔA | *pyrE*(–40) | BW25113_3642 |
| | 928,332 | T > G | *lrp*(N94K) | BW25113_0889 |
| Δ*entC sup*1 | 1,750,786 | A > C | *pykF*(T278P) | BW25113_1676 |
| | 2,374,180 | G > A | *menF*(–58) | BW25113_2265 |
| Δ*entC sup*2 | 2,374,177 | G > A | *menF*(–55) | BW25113_2265 |
| | 4,172,731 | C > T | *rpoB*(P520L) | BW25113_3987 |
| | 2,374,180 | G > A | *menF*(–58) | BW25113_2265 |
| Δ*entC sup*3 | 4,178,500 | C > T | *rpoC*(R1075C) | BW25113_3988 |
| Δ*dgk sup*1 | 4,178,500 | C > T | *rpoC*(R1075C) | BW25113_3988 |
| Δ*dgk sup*2 | 4,172,748 | C > T | *rpoB*(H526Y) | BW25113_3987 |
| Δ*dgk sup*3 | 3,310,740 | A > T | *nusA*(I49N) | BW25113_3169 |

* RNA sequencing was performed on *sup* strains labeled with shaded background.
[a] Mutations in non-coding regions are annotated by the position upstream of the start codon(s) of adjacent genes.
[b] RNA from two independent colonies of the Δ*zwf*-*sup*1 strain was sequenced to verify transcriptional similarity between biological replicates.
IS*5* is a transposable insertion sequence.

**Table 2. Structural features of RNAP corresponding to *sup* mutations.**

| RNAP component | Residue/ substitution (strain) | Region of RNAP structure | Structural features or known mutant phenotypes | References |
|---|---|---|---|---|
| β | D959G (Δ*dapF sup*1) | Lineage-specific βi9 region, adjacent to the β flap | D959 forms a conserved salt bridge with K1032 | [31] |
| | D516G (Δ*zwf sup*3 & *sup*4) | Residues are at the rifampicin binding site, near the DNA/RNA hybrid in the active site | Mutation of D516 is reported to increase RNAP slippage; mutations at D516 and H526 are associated with rifamycin resistance | [32-36] |
| | H526Y (Δ*dgk sup*2) | | | |
| | G713C (Δ*dapF sup*2) | Proximal to the active site | Mutation at nearby residue D711 is associated with increased fitness in stationary phase culture | [37] |
| | P520L (Δ*entC sup*2) | Proximal to the active site | This substitution may change orientation of D516 and H526 | |
| | Q31R (Δ*zwf sup*2) | Residue makes contacts with the H526 helix | Substitution may clash with H526 helix | |
| | D654E (Δ*ppk sup*3) | Proximal to region involved in polymerase interaction with CBR antimicrobials | | [38, 39] |
| | T1037P (*Fast 4a & 4b)* | | | |
| β' | R1075C (Δ*entC sup*3; Δ*dgk sup*1) | β'i6 region, next to the β' jaw region | Mutations in this region can affect transcript cleavage, pausing, and elongation | [5, 31, 40] |
| | R978P/S (Δ*zwf sup*1; Δ*ppk sup*2) | | | |
| | G1207VERG (Δ*ppk sup*1) | β' jaw region | May alter the kinetics of elongation and decrease transcriptional pausing; mutation at this site has been associated with fast growth in defined medium | [5] |
| | E1202K (*Fast 6a & 6b*) | | | |
| | R481H (*Fast 1a*) | Proximal to ppGpp binding site | ppGpp modulates the properties of RNAP in responses to change in nutrients, and can control growth rate. | [41] |
| NusA | I49D (Δ*dgk sup*3) | N-terminal RNAP binding region; binds at RNA exit channel | | [42] |

**Table 3. Mutation combinations with synthetic rescue features**

| Class | Starting Strain | Deleterious Mutation(s) | Rescue Mutation(s) [a] | Growth Media |
|---|---|---|---|---|
| Forward synthetic rescues (S1C Fig) | BW25113 | Δ*dapF* | 4,174,048 A > G; 928,451 C > A | LB |
| | | Δ*dapF* | 4,173,309 G > T; 928,451 C > A | LB |
| | | Δ*zwf* | 4,178,209 C > A; ; 1,301,660 IS*5*(+) | LB |
| Reverse synthetic rescues (S1F Fig) [b] | Δ*dapF-sup*2 | 4,173,309 T > G; 928,451 A > C | Restore *dapF* | M9G |
| | Δ*dgk-sup*1 | 4,178,500 T > C | Restore *dgk* | M9G |
| | Δ*dgk-sup*2 | 4,172,748 T > C | Restore *dgk* | M9G |
| | Δ*entC-sup*3 | 2,374,180 A > G; 4,178,500 T > C | Restore *entC* | M9G |

[a] All rescue mutations are neutral or deleterious in the starting strain background.
[b] Mutations are applied in the reverse order from which they occurred in our experiments.

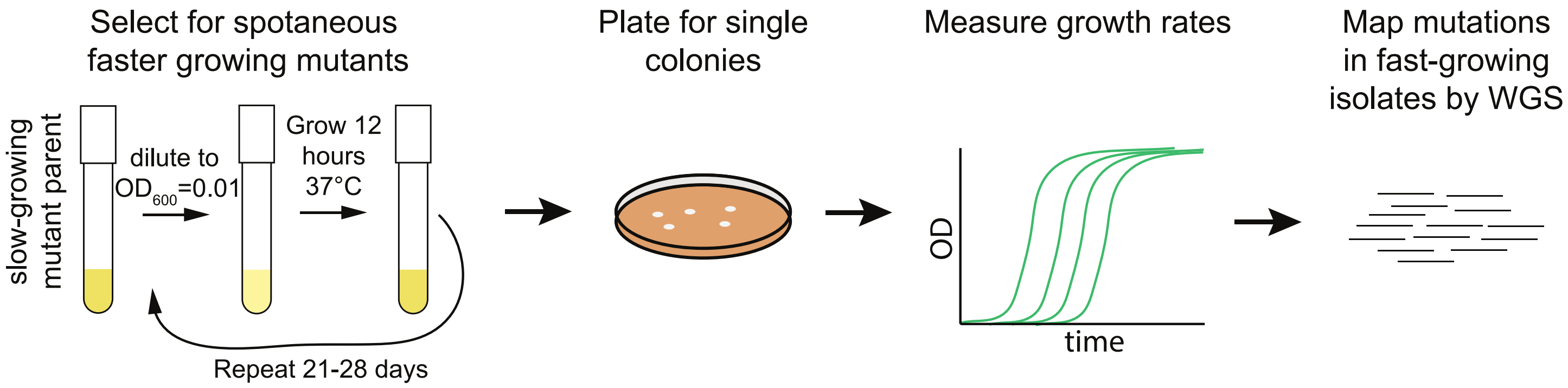

**Fig 1. Experimental approach to isolate and map rescue (*sup*) strains.**

Strains harboring one of five deletions (Δ*zwf*, Δ*dapF*, Δ*dgk*, Δ*entC*, or Δ*ppk*) were suspended in M9 supplemented with 0.4% (w/v) of glucose (M9G) in triplicate. The deletion strains were allowed to grow at 37° C with shaking for 12 hours, at which point they were diluted to OD600 = 0.01. Adaptive evolution proceeded in this manner for 21–28 days (216–611 generations). At the end of this AE period, the culture was plated, colonies were recovered and resuspended in liquid media, and log-phase growth rates were measured, identifying *sup* strains. The genetic lesions associated with rescued growth were identified by whole genome sequencing (WGS).

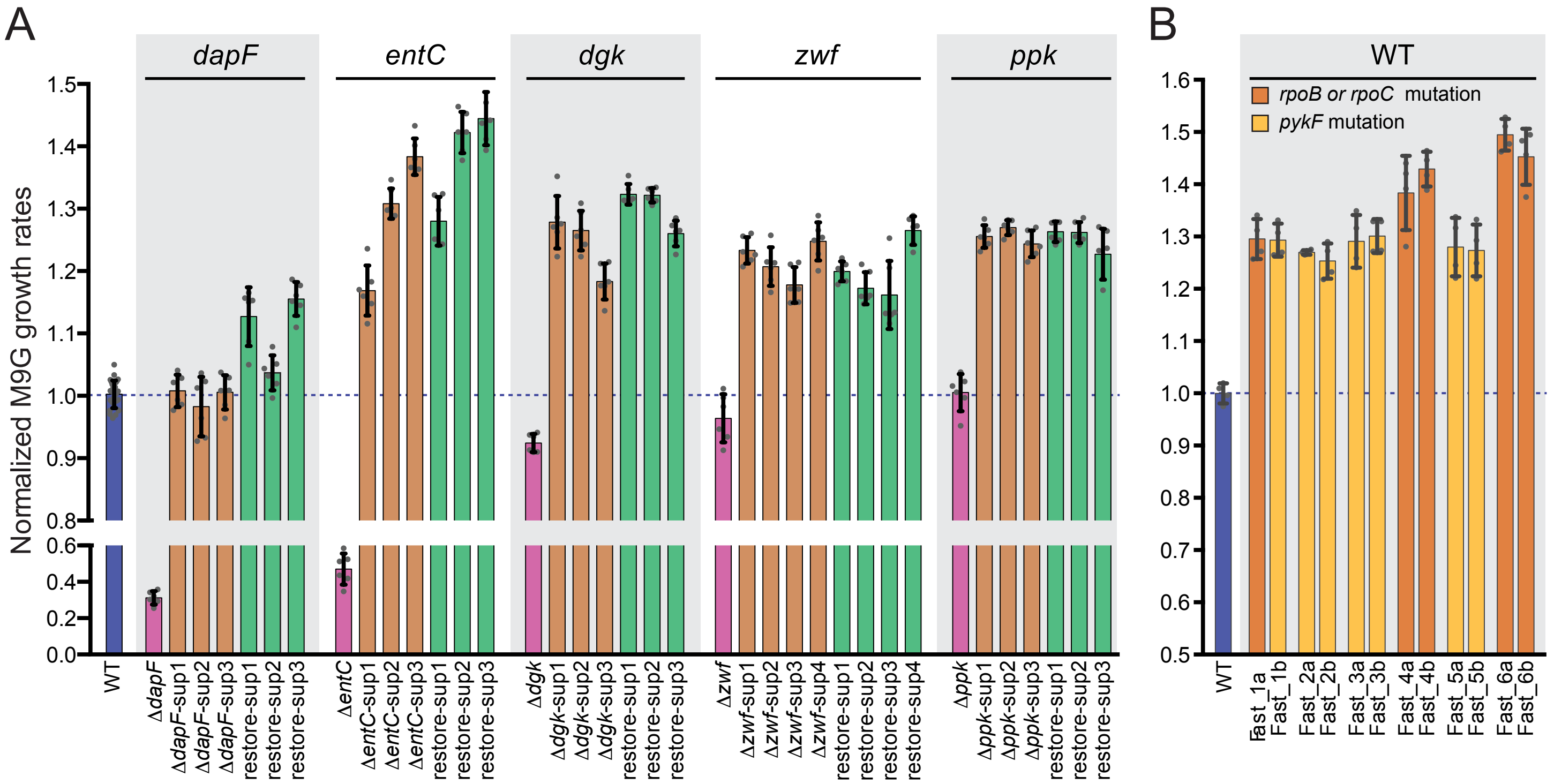

**Fig 2. Growth rates of all E. coli strains assayed in M9G.**

(A) The colored bars indicate the mean growth rates for the WT strain (blue), five primary deletion strains (WT + Δ; pink), 16 fast-growth *sup* strains (WT + Δ + AE; orange), and the same 16 *sup* strains with the respective primary deletion knocked in (WT + AE; green). Each point marks the growth rate of an independent culture (6 replicates for each mutant strain, *sup* strain, and *sup* knock-in strain; 30 replicates for the WT strain). (B) Growth rates of fast-growing adaptively evolved strains derived from a wild-type background presented as in (A). Fast strains are colored according to mutation class: strains with *rpoB* or *rpoC* mutations (dark orange) or with *pykF* mutations (light orange). Points represent 4 independent replicates for each strain. In (A) and (B), the error bars represent the standard deviation over all replicates. All growth rates are normalizedby that of the WT strain in M9G.

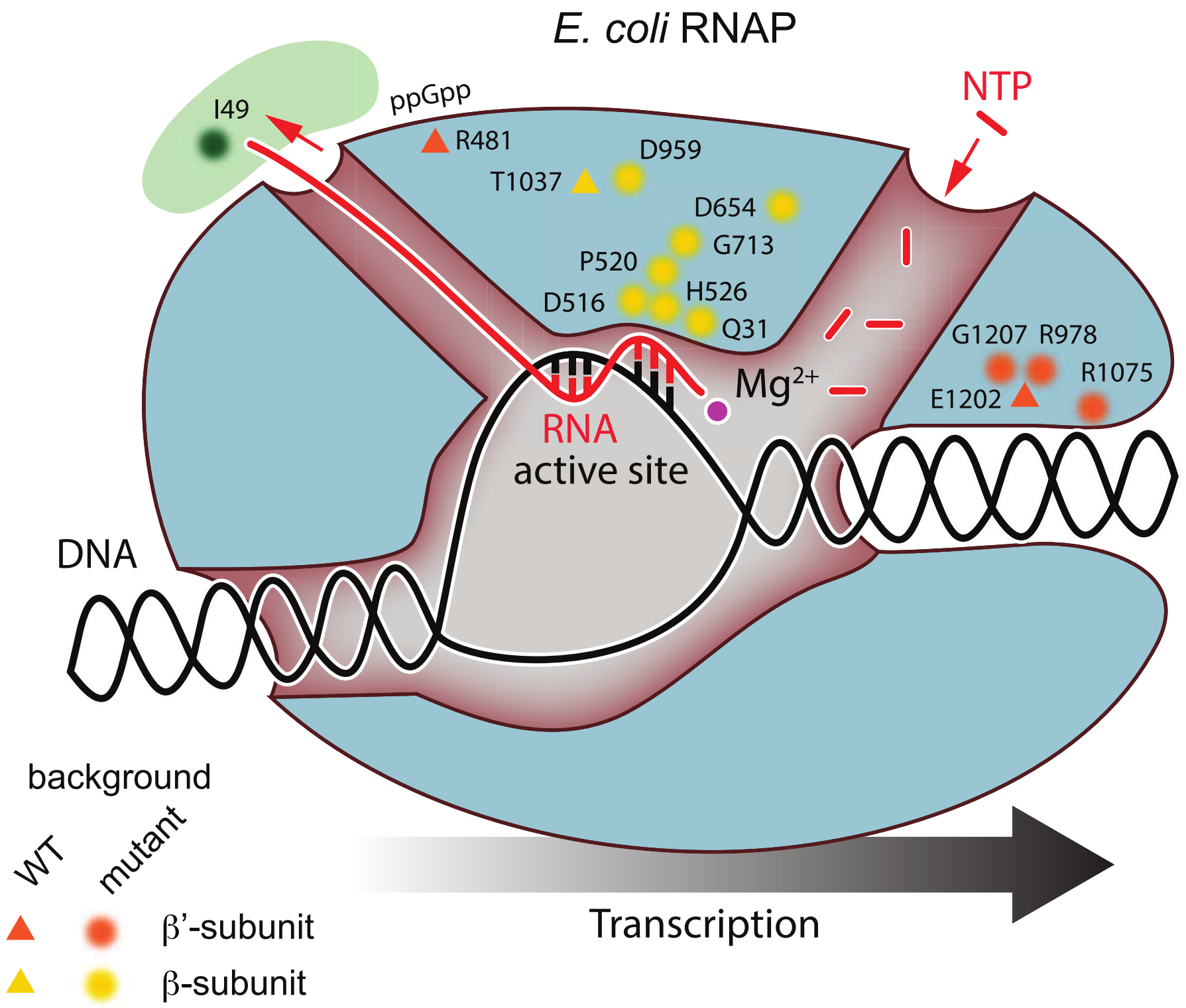

**Fig 3. Model of the *E. coli* RNA polymerase core enzyme complexed with NusA.**

The relative positions of RNAP *sup* mutations identified in our adaptively evolved *sup* strains (selected in M9G), and listed in Table 2 are marked on the polymerase model. *sup* mutations in the β- and β' subunits are color coded yellow and orange, respectively. Polymerase is colored blue; NusA is colored green. DNA is shown in black, and nascent RNA in red. Incoming NTPs are also labeled in red.

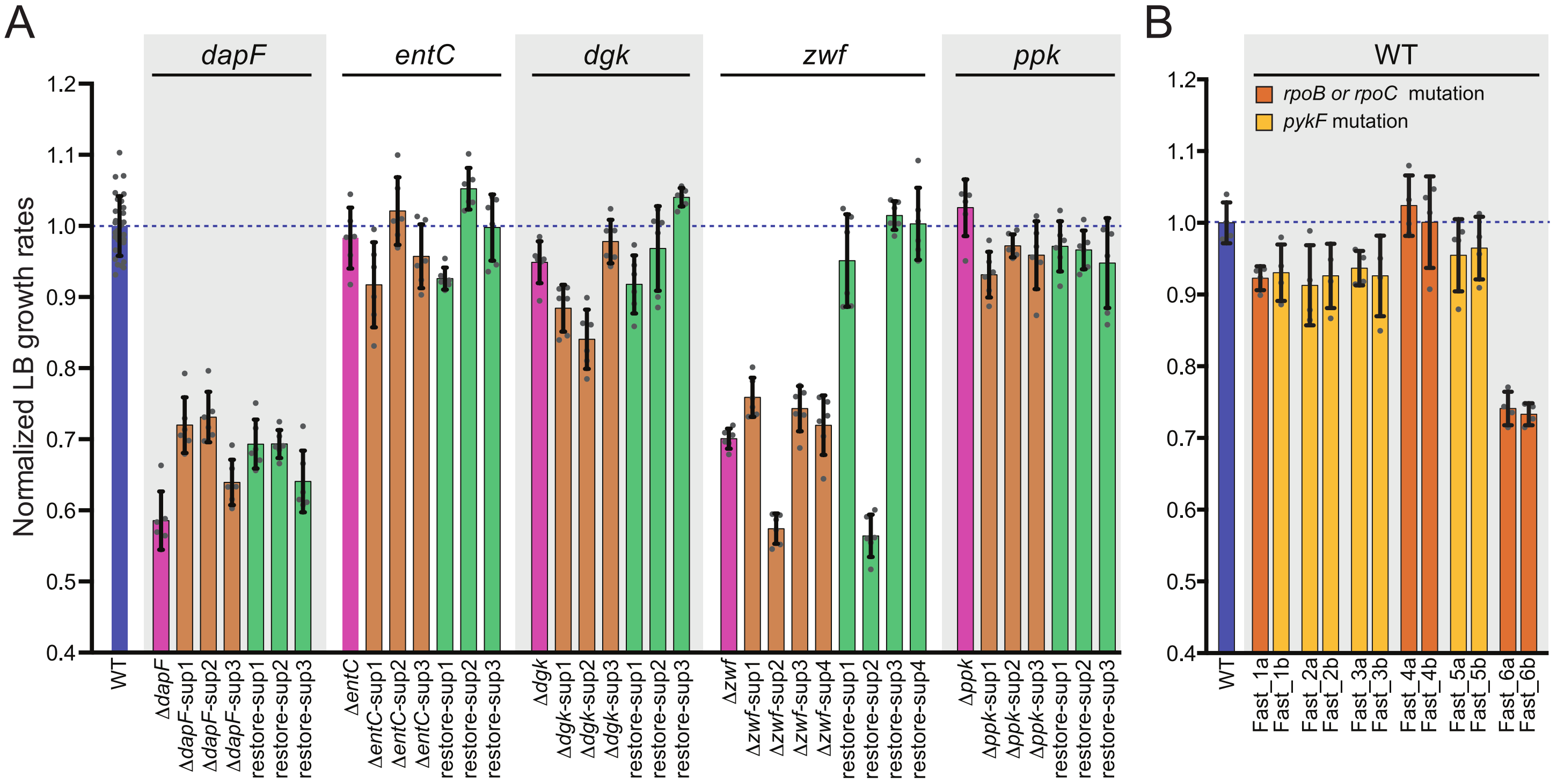

**Fig 4. Growth rates of all *E. coli* strains assayed in LB.**

(A and B) Strains and notation are the same as Fig 2, with the *sup* strains and *Fast* strains selected by AE in M9G before assaying growth in LB. Growth rates are normalized to the WT strain in LB.

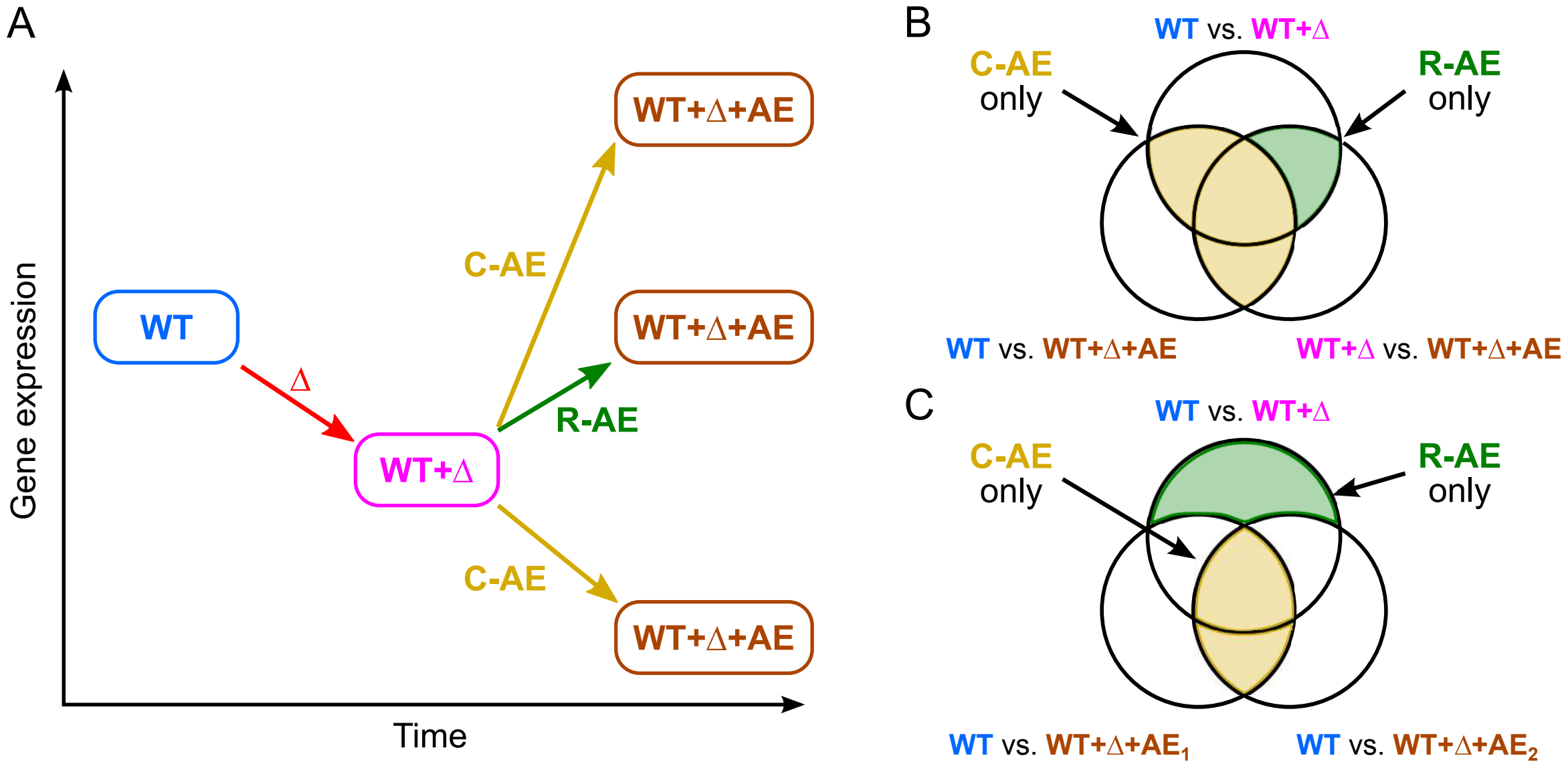

**Fig 5. Restorative versus compensatory adaptive transcriptional responses to a deletion.**

(A) Illustration of the expression of a gene for the WT strain, the deletion strain, and the adaptively evolved (AE) strain for the two possible transcriptional responses: restorative adaptive evolution (R-AE), in which AE restores expression to that of the WT strain, and compensatory adaptive evolution (C-AE), in which the resulting expression is distinct from that of the WT strain. This classification applies to all genes whose expression is significantly perturbed by the gene deletion and/or the AE. (B, C) Venn diagrams indicating the sets of R-AE (green) and C-AE (yellow) genes. The diagram compares gene expression in the WT strain with that of the primary deletion and associated evolved strains, where each circle represents the set of genes exhibiting changes between the pair of strains marked next to it. In (B), all comparisons are for a single run of adaptive evolution and indicate the R-AE and C-AE genes for the resulting *sup* strain. In (C), comparisons are shown for two runs of AE, and the shaded regions mark genes that undergo R-AE and C-AE for both *sup* strains.

A

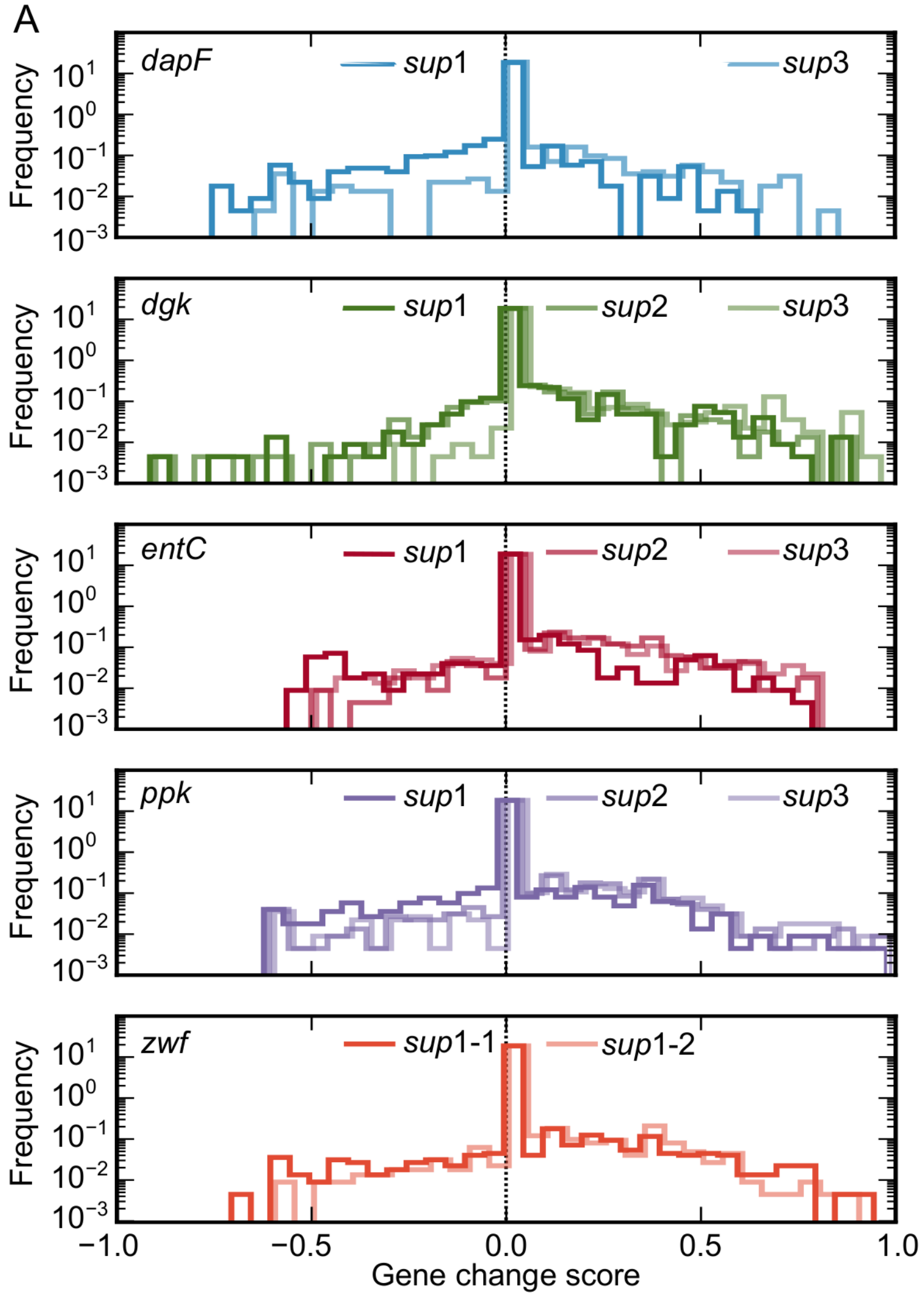

B

| | Compensated | Restored |
|---|---|---|
| Δ*dapF* | *flu, isrC** | *alaE, arrS*, dapF, hdeB*, livH*, livJ*, livK*, lysU, ompT* |
| Δ*dgk* | *dgkA* | *aer, cheA, cheB, cheR, cheW, cheY, cheZ, crfC, csgC, emtA, fadH, flgK*, flgL*, fliC, fliD, fliF, fliG, fliI, fliK, fliS, fliT, fliZ, flxA, hypD, lamB*, malK*, modA*, modB*, modC*, motB, pspD, tap, tar, trg, tsr, uhpT, ves, yafY, ycgR, yhjH, yjcZ, ymdA, ynjH* |
| Δ*entC* | *menF*, ybcW* | *carA*, codA, codB*, gcvT, nadA*, pyrL*, uraA, ydaF* |
| Δ*ppk* | *arrS*, cyaR*, gadA*, gadB*, gadC*, glnK, hdeA*, hdeB*, hdeD*, insA_1, nrdE, nrdH, nrdI, ygcG, yjaA, yphA* | *asnT, asnU, bhsA, dctR, elaD, entH, flgB*, flgC*, flgD*, flgE*, flgG*, fliA, fliF, fliK, fliL, fliN, gtrS, nirB, spf*, tnaA, tomB, ydhC, yfbM, yjgL, ynfO* |
| Δ*zwf* | *dkgA, glnK, tfaX, yeaH, yecF, yjaW, ylcH* | *ariR, arrS*, dctR, gadA*, gadE*, glsA, gsiD mcbA, mdtE, oxc, rzoR, safA, sdaC, slp, spf*, ybgS, ycgZ, yciF, yciG,ydeJ, yegJ, yghX, yhiM, yjdN, yjeV, yjfY, yjjV, ymgA, ymgC yohC, zraP* |

**Fig 6. Evaluation of the compensatory and restorative responses to each gene deletion.**

(A) Histogram of the Gene Change Score (GCS) exhibiting compensatory (negative) and restorative (positive) transcriptional changes for each *sup* strain (marked by different lines). Each case clearly shows a bias toward positive (restorative) changes. (B) List of genes exhibiting the strongest responses (GCS > 0.4 or < –0.2) across all *sup* strains associated with each deletion strain. Genes marked by an asterisk are discussed in more detail in the text.

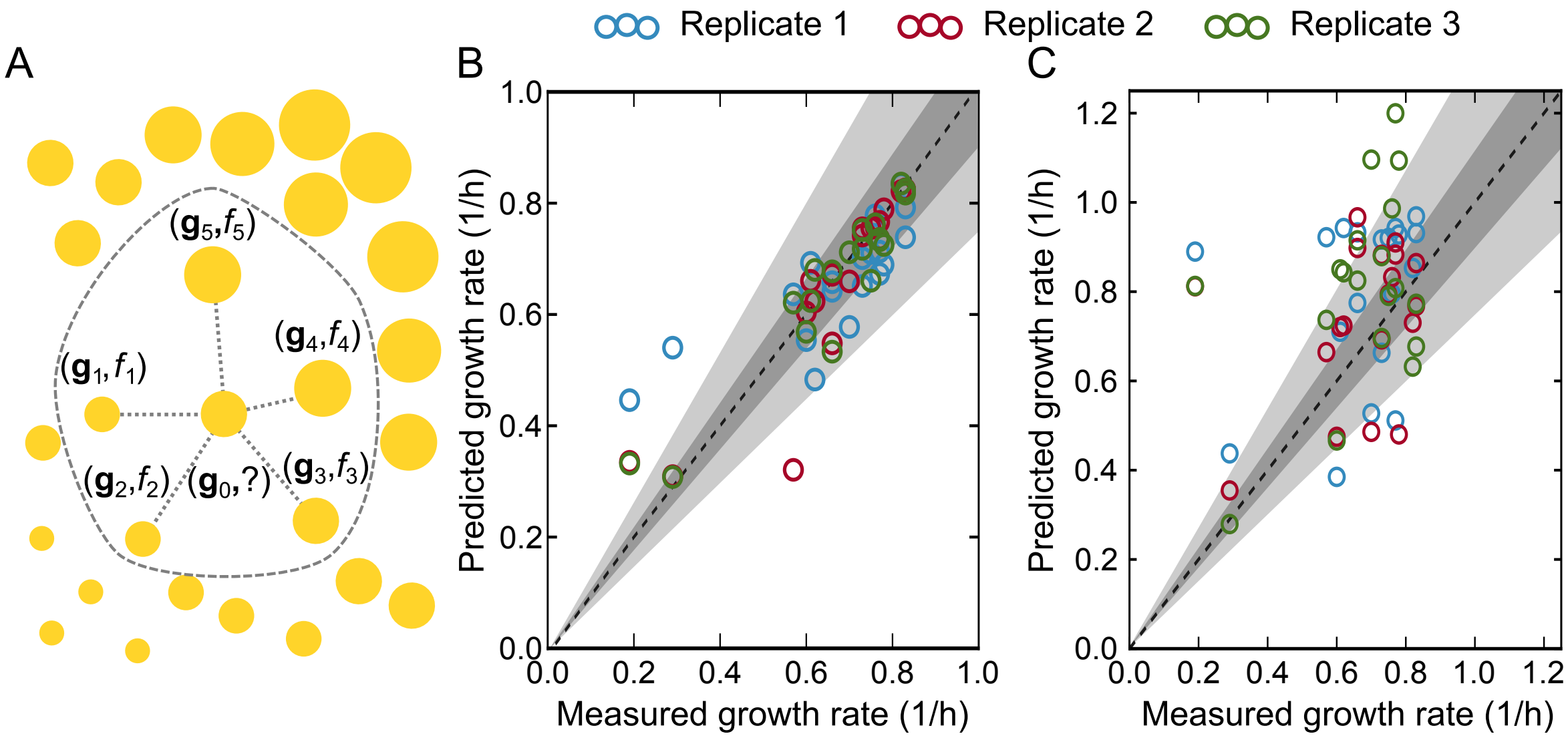

**Fig 7. Description and validation of the *k*-nearest neighbors (KNN) model.**

(A) Illustration of our KNN model for estimating the (unknown, indicated by the question mark) growth rate associated with a given expression profile from the known growth rates for different expression profiles in its neighborhood. Circle centers and sizes indicate gene expression profiles ($\mathbf{g}_i$) and growth rate ($f_i$), respectively. (B) Validation of the KNN model using the gene expression data from all 19 strains for which we measured both gene expression and growth rate in triplicate (underlined strains in Table 1). The KNN analysis was trained on a dataset comprising (i) the data curated in [59], (ii) our transcripomic and growth rate data, and (iii) randomly generated pseudo-profiles (see Materials and Methods for details) for a total of 11,275 datapoints. The dataset was partitioned into fifths that preserved the growth rate distribution and each combination of four fifths was used to train a KNN model that predicted the growth rate of the left-out fifth. For comparison with (C) only predictions regarding our transcriptomic growth data are retained. The figure shows good agreement between the resulting predicted growth rates and the experimentally measured growth rates ($R^2$ = 0.84, *p*-value < $10^{-38}$), where the line of perfect agreement (dashed), 10% error (dark grey) and 25% error (light grey) are included as a reference. (C) Prediction of our measured growth rates by the KNN analysis when trained on publicly available data only. Results are presented for Dataset D2 in S4 Table ($R^2$ = 0.11, *p*-value < 0.015). Colors and dashed line have the same meaning as in (B). The accuracy in (C) is lower than in (B) because, with only a limited number of experiments to use for training, outlier transcriptional states remain under-sampled making it hard to resolve the growth rate of these states.

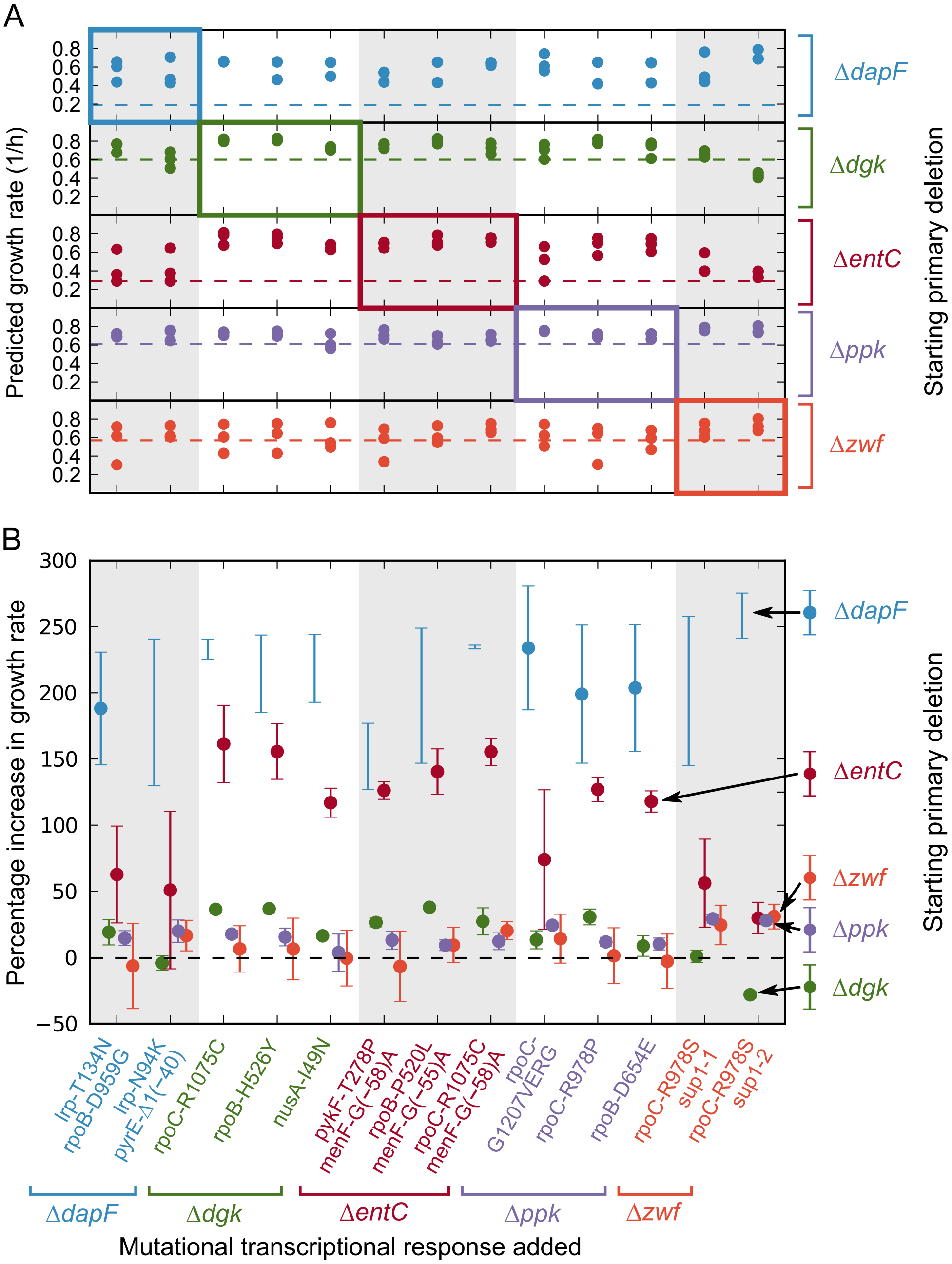

**Fig 8. Predicted growth effects of sup mutations in all alternative deletion backgrounds.**

(A) Predicted growth rates for expression profiles constructed by adding each of $n$ = 3 replicate of the primary deletion (right color code) to each mutational transcriptional response (color coded at the bottom by the primary deletion). Dashed lines indicate the measured growth rates of the deletion strains. Note that points in the highlighted diagonal blocks lie above the dashed lines indicating the consistency with which mutations rescue growth in the primary deletion background in which they were selected. (B) Mean predicted percent increase over the growth rate of the measured deletion strain (right color code) when the mutational transcriptional response (bottom color code) is added. The dashed line demarcates the boundary between beneficial and deleterious mutational effects. Error bars standard error of the mean over the replicates, and each starting strain is shifted horizontally by the width of a dot for visibility.

## Supplemental Information

## Tables and Figures

**S1 Table. Specific growth rate and genetic selection information for *E. coli* strains grown on M9G.** Data are reported for *E. coli* BW25113 (WT); five primary mutant strains (Δ*zwf*, Δ*ppk*, Δ*dapF*, Δ*entC*, and Δ*dgk*); 16 independent *sup* strains (which rescue slow growth of metabolic mutants); 16 independent knock-in strains in which the primary mutation was restored to the WT allele (noted in table as 'restore'); and 12 *Fast* strains evolved from WT paired with a repeat of wild-type from paired growth experiments.

| Strain | Mean population growth rate (1/h ± SD) | Estimated total doublings (range) | Final fitness (WT = 1.00) | Number of serial transfers |
|---|---|---|---|---|
| *E. coli* K-12 BW25113 | 0.62 ± 0.03 | NA | 1.00 | NA |
| *E. coli* Δ*zwf* | 0.57 ± 0.03 | NA | 0.91 | NA |
| *E. coli* Δ*zwf sup*1 | 0.73 ± 0.01 | 265–510 | 1.18 | 56 |
| *E. coli* Δ*zwf sup*2 | 0.72 ± 0.02 | 265–341 | 1.16 | 56 |
| *E. coli* Δ*zwf sup*3 | 0.70 ± 0.02 | 265–498 | 1.13 | 56 |
| *E. coli* Δ*zwf sup*4 | 0.74 ± 0.02 | 265–497 | 1.20 | 56 |
| *zwf restore-sup*1 | 0.71 ± 0.01 | NA | 1.15 | NA |
| *zwf restore-sup*2 | 0.70 ± 0.02 | NA | 1.12 | NA |
| *zwf restore-sup*3 | 0.69 ± 0.03 | NA | 1.11 | NA |
| *zwf restore-sup*4 | 0.75 ± 0.01 | NA | 1.21 | NA |
| *E. coli* Δ*ppk* | 0.61 ± 0.02 | NA | 0.99 | NA |
| *E. coli* Δ*ppk sup*1 | 0.76 ± 0.01 | 334–590 | 1.23 | 56 |
| *E. coli* Δ*ppk sup*2 | 0.77 ± 0.01 | 334–586 | 1.24 | 56 |
| *E. coli* Δ*ppk sup*3 | 0.75 ± 0.01 | 334–562 | 1.21 | 56 |
| *ppk restore-sup*1 | 0.76 ± 0.01 | NA | 1.23 | NA |
| *ppk restore-sup*2 | 0.76 ± 0.01 | NA | 1.23 | NA |
| *ppk restore-sup*3 | 0.74 ± 0.02 | NA | 1.20 | NA |
| *E. coli* Δ*dapF* | 0.19 ± 0.03 | NA | 0.31 | NA |
| *E. coli* Δ*dapF sup*1 | 0.66 ± 0.02 | 216–482 | 1.06 | 56 |
| *E. coli* Δ*dapF sup*2 | 0.64 ± 0.03 | 216–494 | 1.04 | 56 |
| *E. coli* Δ*dapF sup*3 | 0.66 ± 0.02 | 216–497 | 1.06 | 56 |
| *dapF restore-sup*1 | 0.74 ± 0.03 | NA | 1.19 | NA |
| *dapF restore-sup*2 | 0.68 ± 0.02 | NA | 1.10 | NA |
| *dapF restore-sup*3 | 0.76 ± 0.02 | NA | 1.22 | NA |
| *E. coli* Δ*entC* | 0.29 ± 0.04 | NA | 0.47 | NA |
| *E. coli* Δ*entC sup*1 | 0.70 ± 0.02 | 232–478 | 1.13 | 42 |
| *E. coli* Δ*entC sup*2 | 0.78 ± 0.01 | 232–534 | 1.26 | 42 |
| *E. coli* Δ*entC sup*3 | 0.83 ± 0.02 | 232–468 | 1.33 | 42 |
| *entC restore-sup*1 | 0.76 ± 0.02 | NA | 1.23 | NA |
| *entC restore-sup*2 | 0.85 ± 0.02 | NA | 1.37 | NA |
| *entC restore-sup*3 | 0.86 ± 0.02 | NA | 1.39 | NA |
| *E. coli* Δ*dgk* | 0.60 ± 0.01 | NA | 0.96 | NA |
| *E. coli* Δ*dgk sup*1 | 0.83 ± 0.03 | 452–594 | 1.34 | 42 |
| *E. coli* Δ*dgk sup*2 | 0.82 ± 0.02 | 452–611 | 1.32 | 42 |
| *E. coli* Δ*dgk sup*3 | 0.77 ± 0.02 | 452–587 | 1.24 | 42 |
| *dgk restore-sup*1 | 0.86 ± 0.01 | NA | 1.38 | NA |
| *dgk restore-sup*2 | 0.86 ± 0.01 | NA | 1.38 | NA |
| *dgk restore-sup*3 | 0.82 ± 0.01 | NA | 1.32 | NA |
| | | | | |
| *E. coli* K-12 BW25113 | 0.60 ± 0.01 | NA | 1.00 | NA |
| Fast_1a | 0.78 ± 0.02 | 160-240 | 1.30 | 24 |
| Fast_1b | 0.78 ± 0.02 | 160-240 | 1.29 | 24 |
| Fast_2a | 0.76 ± 0.003 | 160-240 | 1.27 | 24 |
| Fast_2b | 0.75 ± 0.02 | 160-240 | 1.25 | 24 |
| Fast_3a | 0.77 ± 0.03 | 160-240 | 1.29 | 24 |
| Fast_3b | 0.78 ± 0.02 | 160-240 | 1.30 | 24 |
| Fast_4a | 0.83 ± 0.04 | 160-240 | 1.38 | 24 |
| Fast_4b | 0.86 ± 0.02 | 160-240 | 1.43 | 24 |
| Fast_5a | 0.77 ± 0.03 | 160-240 | 1.28 | 24 |
| Fast_5b | 0.76 ± 0.03 | 160-240 | 1.27 | 24 |
| Fast_6a | 0.90 ± 0.02 | 160-240 | 1.49 | 24 |
| Fast_6b | 0.87 ± 0.03 | 160-240 | 1.45 | 24 |

**S2 Table. LB-cultivated growth rate data.** Specific growth rate (in 1/h) is reported for the strains enumerated in S1 Table cultivated in LB.

| Strain | Mean population growth rate (1/h ± SD) | Final fitness (wild type = 1.00) |
|---|---|---|
| *E. coli* K-12 BW25113 | 1.64 ± 0.08 | 1.00 |
| *E. coli* Δ*zwf* | 1.12 ± 0.02 | 0.68 |
| *E. coli* Δ*zwf sup*1 | 1.21 ± 0.04 | 0.74 |
| *E. coli* Δ*zwf sup*2 | 0.92 ± 0.03 | 0.56 |
| *E. coli* Δ*zwf sup*3 | 1.18 ± 0.05 | 0.72 |
| *E. coli* Δ*zwf sup*4 | 1.15 ± 0.07 | 0.70 |
| *zwf restore-sup*1 | 1.52 ± 0.10 | 0.93 |
| *zwf restore-sup*2 | 0.90 ± 0.05 | 0.55 |
| *zwf restore-sup*3 | 1.62 ± 0.03 | 0.99 |
| *zwf restore-sup*4 | 1.60 ± 0.08 | 0.98 |
| *E. coli* Δ*ppk* | 1.64 ± 0.06 | 1.00 |
| *E. coli* Δ*ppk sup*1 | 1.49 ± 0.05 | 0.91 |
| *E. coli* Δ*ppk sup*2 | 1.55 ± 0.03 | 0.95 |
| *E. coli* Δ*ppk sup*3 | 1.53 ± 0.08 | 0.93 |
| *ppk restore-sup*1 | 1.55 ± 0.06 | 0.95 |
| *ppk restore-sup*2 | 1.54 ± 0.04 | 0.94 |
| *ppk restore-sup*3 | 1.51 ± 0.10 | 0.92 |
| *E. coli* Δ*dapF* | 0.95 ± 0.07 | 0.58 |
| *E. coli* Δ*dapF sup*1 | 1.17 ± 0.06 | 0.71 |
| *E. coli* Δ*dapF sup*2 | 1.19 ± 0.06 | 0.73 |
| *E. coli* Δ*dapF sup*3 | 1.04 ± 0.05 | 0.63 |
| *dapF restore-sup*1 | 1.13 ± 0.06 | 0.69 |
| *dapF restore-sup*2 | 1.13 ± 0.03 | 0.69 |
| *dapF restore-sup*3 | 1.04 ± 0.07 | 0.63 |
| *E. coli* Δ*entC* | 1.67 ± 0.07 | 1.02 |
| *E. coli* Δ*entC sup*1 | 1.56 ± 0.10 | 0.95 |
| *E. coli* Δ*entC sup*2 | 1.74 ± 0.08 | 1.06 |
| *E. coli* Δ*entC sup*3 | 1.63 ± 0.08 | 0.99 |
| *entC restore-sup*1 | 1.58 ± 0.03 | 0.96 |
| *entC restore-sup*2 | 1.79 ± 0.05 | 1.09 |
| *entC restore-sup*3 | 1.70 ± 0.08 | 1.04 |
| *E. coli* Δ*dgk* | 1.61 ± 0.05 | 0.98 |
| *E. coli* Δ*dgk sup*1 | 1.50 ± 0.06 | 0.91 |
| *E. coli* Δ*dgk sup*2 | 1.43 ± 0.07 | 0.87 |
| *E. coli* Δ*dgk sup*3 | 1.66 ± 0.05 | 1.01 |
| *dgk restore-sup*1 | 1.56 ± 0.07 | 0.95 |
| *dgk restore-sup*2 | 1.64 ± 0.10 | 1.00 |
| *dgk restore-sup*3 | 1.77 ± 0.02 | 1.08 |
| | | |
| *E. coli* K-12 BW25113 | 1.85 ± 0.05 | 1.00 |
| Fast_1a | 1.70 ± 0.03 | 0.92 |
| Fast_1b | 1.72 ± 0.07 | 0.93 |
| Fast_2a | 1.68 ± 0.10 | 0.91 |
| Fast_2b | 1.71 ± 0.08 | 0.93 |
| Fast_3a | 1.73 ± 0.04 | 0.94 |
| Fast_3b | 1.71 ± 0.10 | 0.93 |
| Fast_4a | 1.89 ± 0.08 | 1.02 |
| Fast_4b | 1.85 ± 0.12 | 1.00 |
| Fast_5a | 1.76 ± 0.09 | 0.95 |
| Fast_5b | 1.78 ± 0.08 | 0.96 |
| Fast_6a | 1.37 ± 0.04 | 0.74 |
| Fast_6b | 1.35 ± 0.03 | 0.73 |

**S3 Table. List of RNA-seq studies in *E. coli* Δ*hns* or Δ*rpoS* in that were used in S2 Fig.** These data were obtained by searching the NCBI SRA database for "*hns*" and "*rpoS*" on January 30, 2018. Reads were aligned to *E. coli* MG1655 (NC_000913) using Rockhopper [98]. The list presents only experiments that successfully aligned reads to the genome. The "Run Number" corresponds to the sequencing run number in the NCBI Sequencing Read Archive (SRA) database. The "Figure Label" is the label in S2 Fig associated with each run. The "Project Number" indicates whether runs were part of the same project in the SRA database. The GEO Accession is the cross-referenced identifier in the GEO database. The WT Runs are the sequencing runs against which fold change was measured. Their associated GEO Accessions are in the WT GEO Accession column.

| Run Number | Figure Label | Project No. | GEO GSM ID | WT Runs | WT GEO ID |
|---|---|---|---|---|---|
| ERR1450560 | MG1655 Δ*hns* 1 in LB | ERP016032 | | ERR1450564, ERR1450566 | |
| ERR1450562 | MG1655 Δ*hns* 1 in LB | ERP016032 | | ERR1450564, ERR1450566 | |
| SRR1449308 | MG1655 Δ*hns* Δ*stpA* 1 in LB | SRP043518 | | SRR1449306, SRR1449307 | |
| SRR1449309 | MG1655 Δ*hns* Δ*stpA* 1 in LB | SRP043518 | | SRR1449306, SRR1449307 | |
| SRR2637697 | EDL933 Δ*hns* in LB | SRP064749 | GSM1906889 | SRR2637695, SRR2637696 | GSM1906887, GSM1906888 |
| SRR2637698 | EDL933 Δ*hns* in LB | SRP064749 | GSM1906890 | SRR2637695, SRR2637696 | GSM1906887, GSM1906888 |
| SRR546799 | MG1655 Δ*hns* 2 in LB | SRP015118 | GSM991202 | SRR546813, SRR546814 | GSM991216, GSM991217 |
| SRR546800 | MG1655 Δ*hns* 2 in LB | SRP015118 | GSM991203 | SRR546813, SRR546814 | GSM991216, GSM991217 |
| SRR546807 | MG1655 Δ*hns* Δ*stpA* 2 in LB | SRP015118 | GSM991210 | SRR546813, SRR546814 | GSM991216, GSM991217 |
| SRR546808 | MG1655 Δ*hns* Δ*stpA* 2 in LB | SRP015118 | GSM991211 | SRR546813, SRR546814 | GSM991216, GSM991217 |
| SRR2932645 | MG1655 Δ*rpoS* EE | SRP065958 | GSM1933962 | SRR2932637, SRR2932638 | GSM1933954, GSM1933955 |
| SRR2932646 | MG1655 Δ*rpoS* EE | SRP065958 | GSM1933963 | SRR2932637, SRR2932638 | GSM1933954, GSM1933955 |
| SRR2932655 | MG1655 Δ*rpoS* ME | SRP065958 | GSM1933972 | SRR2932647, SRR2932648 | GSM1933964, GSM1933965 |
| SRR2932656 | MG1655 Δ*rpoS* ME | SRP065958 | GSM1933973 | SRR2932647, SRR2932648 | GSM1933964, GSM1933965 |
| SRR2932665 | MG1655 Δ*rpoS* TS | SRP065958 | GSM1933982 | SRR2932657, SRR2932658 | GSM1933974, GSM1933975 |
| SRR2932666 | MG1655 Δ*rpoS* TS | SRP065958 | GSM1933983 | SRR2932657, SRR2932658 | GSM1933974, GSM1933975 |
| SRR2932675 | MG1655 Δ*rpoS* S | SRP065958 | GSM1933992 | SRR2932667, SRR2932668 | GSM1933984, GSM1933985 |
| SRR2932676 | MG1655 Δ*rpoS* S | SRP065958 | GSM1933993 | SRR2932667, SRR2932668 | GSM1933984, GSM1933985 |
| SRR2932685 | MG1655 Δ*rpoS* LS | SRP065958 | GSM1934002 | SRR2932677, SRR2932678 | GSM1933994, GSM1933995 |
| SRR2932686 | MG1655 Δ*rpoS* LS | SRP065958 | GSM1934003 | SRR2932677, SRR2932678 | GSM1933994, GSM1933995 |
| SRR4416204 | BW27786 26% *rpoS* in LB | SRP091400 | GSM2341963 | SRR4416202, SRR4416205 | GSM2341961, GSM2341964 |
| SRR4416207 | BW27786 26% *rpoS* in LB | SRP091400 | GSM2341966 | SRR4416202, SRR4416205 | GSM2341961, GSM2341964 |
| SRR4416203 | BW27786 Δ*rpoS* in LB | SRP091400 | GSM2341962 | SRR4416202, SRR4416205 | GSM2341961, GSM2341964 |
| SRR4416206 | BW27786 Δ*rpoS* in LB | SRP091400 | GSM2341965 | SRR4416202, SRR4416205 | GSM2341961, GSM2341964 |

**S4 Table. Description of datasets used in the out-of-sample KNN analysis and prediction performance of our measured growth rates using our RNA-Seq data as input.** Tab "Training Data Description" defines the GEO Accession numbers composing each dataset listed in the column titled "Set Name". Tab "Out of Sample Performance" lists summary predictive performance as measured by the coefficient of determination ($R^2$) and the number of experiments predicted within either 25% or 10% accuracy. The size of the dataset used to measure correlations and train the model is detailed in the column "No. Experiments (with Growth)". The final column lists the number of experiments predicted within the specified accuracy under the leave-one-strain-out model.

| **Dataset** | **Ref. 59** | **D1** | **D2** | **D3** |
|---|---|---|---|---|
| Added GEO Series Accession Numbers | Ref. 59 | Ref. 59 & {GSE5329, GSE59377, GSE61327, GSE97944} | S1 & {GSE28412, GSE51581, GSE82343} | S1 & {GSE28412, GSE51581, GSE49296, GSE59050, GSE78756} |
| $R^2$ | 0.06 | 0.09 | 0.11 | 0.13 |
| No. Experiments (w/ growth) | 2,198 (589) | 2,266 (654) | 2,302 (690) | 2,395 (783) |
| No. Shared Genes | 4,189 | 4,074 | 3,608 | 1,683 |
| No. Features | 3 | 12 | 9 | 10 |
| No. Experiments predicted within 25% (10%) | 0 (0) | 30 (15) | 35 (13) | 0 (2) |
| No. Experiments predicted within 25% (10%) when using data from the present study, excluding replicates | 47 (20) | 46 (28) | 45 (28) | 49 (28) |

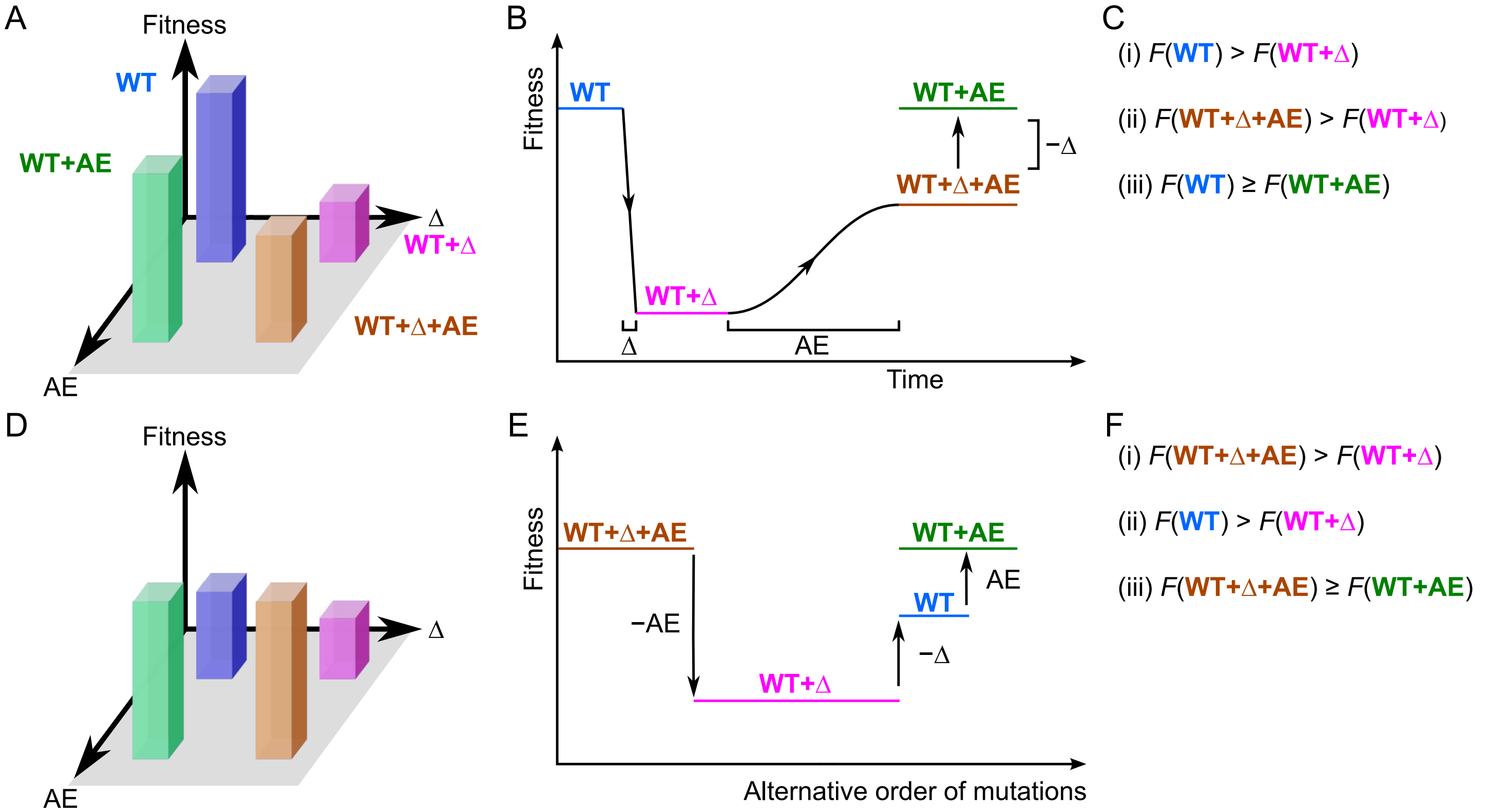

**S1 Fig. Experimental approach to identify synthetic rescue interactions.**

(A) Bar graph of fitness (i.e., growth rate) illustrating synthetic rescue epistasis in which a gene deletion (Δ) is rescued by adaptively evolved *sup* mutations. The fitness of the WT strain (blue) decreases upon gene deletion (WT + Δ, pink) and increases upon acquisition of *sup* mutations during adaptive evolution (AE) (WT + Δ + AE, orange). The impact of *sup* mutations acquired during AE on the WT strain (WT + AE, green) is neutral at best. (B) Order of the mutations in (A), where the WT + AE strain is constructed by restoring the primary gene deletion in the *sup* strain. (C) Mathematical conditions defining synthetic rescues, where F is the fitness of each genetic background. (D, E) *A posteriori* approach to identify synthetic rescue interactions, considering an alternative order of mutations. The initial strain is WT + Δ + AE; genetic perturbations are the restoration of the *sup* mutations acquired through AE (–AE) and the restoration of the primary gene deletion mutation (–Δ). (F) Counterpart of (C) for the alternative order in (D, E). Specific examples of synthetic rescues from our experiments are presented in Table 3 for both mutational orders.

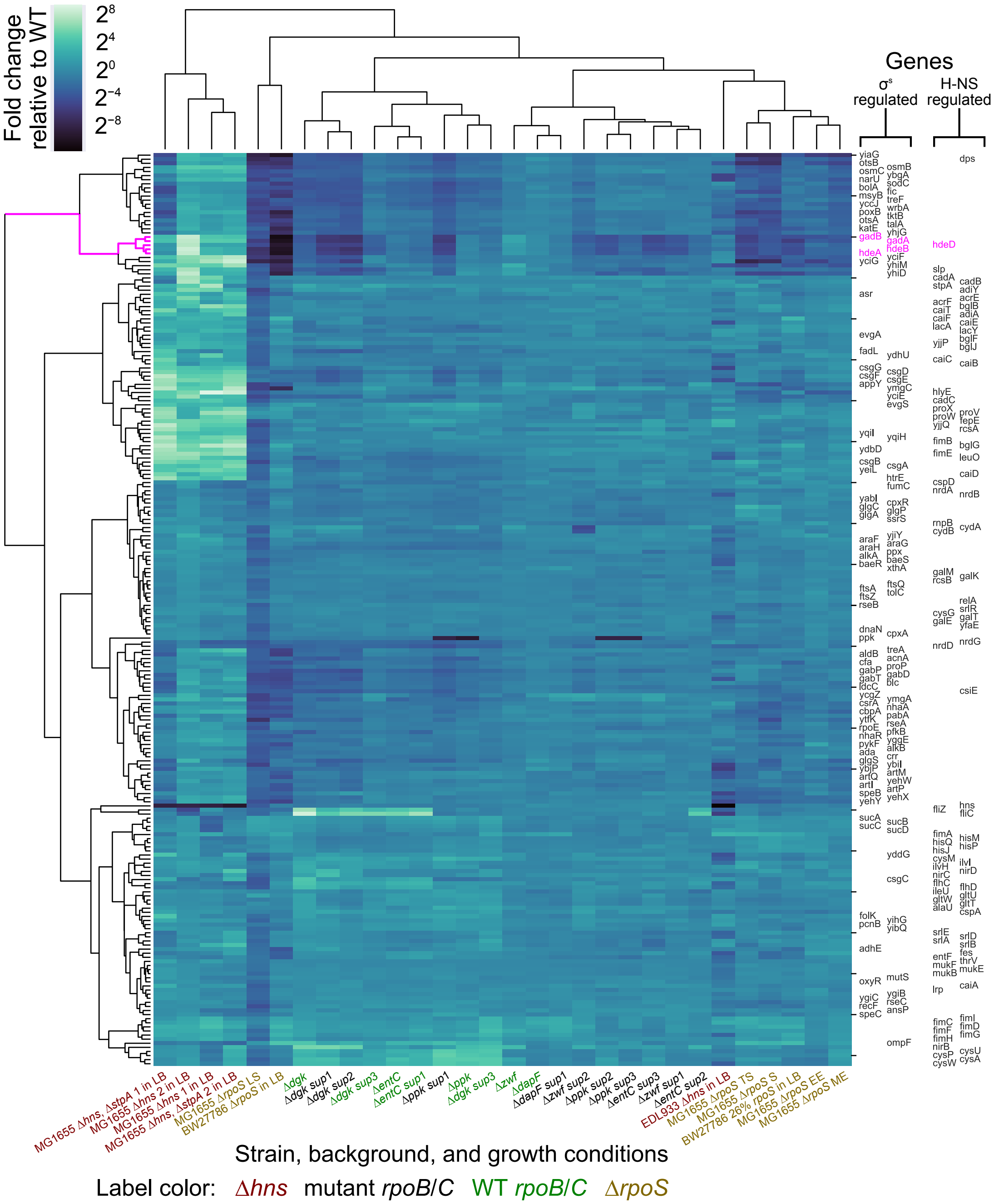

**S2 Fig. Heatmap of RNA-Seq data for genes regulated by H-NS and $\sigma^s$.**

Rows and columns correspond to genes and samples, respectively. Genes regulated by H-NS and $\sigma^s$ were selected based on [49], and filtered to remove ribosomal RNA genes and genes absent in strain BW25113. In particular, the *hde* and *gad* genes discussed in the text are highlighted in magenta. RNA-sequencing experiments obtained from the NCBI SRA database are listed in S4 Table. Labels detailing strain and growth condition are color-coded for strains with Δ*hns* (red), Δ*rpoS* (gold), unmutated *rpoBC* (green), and mutant *rpoBC* (black). If not indicated otherwise, the strain genetic background is K12 BW25113 and the growth condition is exponential phase in M9. The dendrograms indicate the relatedness of the transcriptional profiles as measured by the Ward metric. Transcriptional fold changes were measured against their corresponding WT sequencing runs, as indicated in S4 Table. Growth phase abbreviations: EE – Early Exponential; ME – Mid-Exponential; TS – Transition to Stationary; S – Stationary; LS – Late Stationary.